\documentclass{article}

\usepackage[spanish]{babel}
\usepackage{verbatim}
\usepackage{graphicx}
\usepackage{hyperref}
\usepackage{amsmath}
\usepackage{amsfonts}

\usepackage[utf8]{inputenc}
\usepackage{tikz}
\usetikzlibrary{calc,matrix,fit,backgrounds}
\newcommand{\Zset}{\mathbb{Z}}

\title{Procesamiento topo-geométrico de imágenes neuronales}
\author{Ana Romero, Jónathan Heras, Gadea Mata, Miguel Morales y Julio Rubio} 
\date{}

\begin{document}

\maketitle

 \begin{abstract}
 El lector habitual de \textit{La Gaceta} tal vez recuerde, al contemplar algunas de las im\'agenes de este art{\'\i}culo --como la Figura 2--, la portada de esta misma revista, en su n\'umero 2 del a\~no 2013. En efecto, tanto la imagen de aquella portada como el art{\'\i}culo que el lector tiene entre sus manos son obra del mismo grupo de investigadores, matem\'aticos, inform\'aticos y bi\'ologos, congregados en torno al proyecto \textit{SpineUp}, ver  \url{http://spineup.es/}. Queremos agradecer a todos ellos  la atenci\'on que han tenido al  responder, con prontitud y generosidad,  a nuestra invitaci\'on para contribuir a esta Columna de Matem\'atica Computacional.   Estamos seguros  de que el lector disfrutar\'a con la atractiva y detallada descripci\'on del singular proyecto que este grupo est\'a desarrollando en el \'ambito de las aplicaciones de la topolog{\'\i}a algebraica --más precisamente, del c\'alculo simb\'olico en el contexto de la topolog{\'\i}a algebraica-- al estudio de las enfermedades neurodegenerativas, como el Alzh\'eimer.

Una \'ultima reflexi\'on. El trabajo en estos campos a caballo entre disciplinas muy diversas es,  como el lector apreciar\'a, extraordinariamente sugerente y reconfortante para el intelecto.  Pero nuestra valoraci\'on de esta tarea debe ir más all\'a de la admiraci\'on intelectual y de la utilizaci\'on, más o menos interesada, como  propaganda, de este y de otros,  pocos,  ejemplos que conozcamos de grupos matem\'aticos  cuya  investigaci\'on se desarrolla desde la colaboraci\'on con equipos de otros \'ambitos cient{\'\i}ficos o tecnol\'ogicos. Esperemos que art{\'\i}culos como este contribuyan a la necesaria sensibilizaci\'on social para que nuestras autoridades cient{\'\i}ficas valoren adecuadamente las dificultades especiales que conlleva este tipo de trabajo en colaboraci\'on y lo apoyen decididamente.
 \end{abstract}



\section{Introducci\'on}

El texto que sigue a continuación podría ser presentado de diferentes formas. Podría considerarse
una historia de \emph{amistad}. O más concretamente de \emph{amistad científica}, o de
\emph{buen entendimiento científico}: el que surgió al converger en La Rioja, en el año 2010,
el equipo de biología sobre \emph{Plasticidad Sináptica Estructural} (PSE, en el resto del artículo), coordinado por el cuarto autor de este artículo (Miguel Morales), y el equipo
de \emph{Programación y Cálculo Simbólico}, coordinado por el quinto autor (Julio Rubio). Hubo
una sintonía inmediata, pese a los diferentes lenguajes empleados por unos y otros (aspecto que
será destacado más adelante), y una coincidencia en el impulso por investigar en temas nuevos,
que emergían de dos campos distantes.

También podría ser relatada como una historia de \emph{buena suerte}, al coincidir en el tiempo
(¡y en el espacio!), la necesidad de encontrar aplicaciones biomédicas dentro del proyecto europeo
ForMath \cite{Formath} (dedicado a la formalización de las Matemáticas) y la necesidad en el grupo PSE de realizar experimentos masivos para facilitar
la explotación de ciertos hallazgos que acababan de patentar.

Incluso podría presentarse como una  \emph{fábula política}. Sin embargo, hemos optado por una
introducción más clásica, basada en la noción de \emph{reducción científica}, en la que, por un
proceso de especialización, unas disciplinas se apoyan en otras, hasta obtener unos resultados
que atraviesan todas las capas científicas recorridas.

Así, comenzamos con un problema \emph{social}, como es el aumento en nuestra sociedad de las
enfermedades neurodegenerativas (siendo la enfermedad de Alzhéimer una de las más extendidas).
Este problema social puede ser \emph{reducido} a un problema \emph{médico}, que requiere de métodos de diagnóstico y de mecanismos farmacológicos de tratamiento. A su vez, el problema médico puede ser
\emph{reducido} a un problema \emph{biológico}, en el que el objeto de estudio ha cambiado, entre
otras posibilidades, a la estructura neuronal del cerebro humano. Hay que dejar claro que estas reducciones no pueden ser  consideradas \emph{completas}; al igual que no es realista, ni razonable, considerar que el punto de vista médico agota el tratamiento social de la enfermedad, tampoco lo será independizar unas
disciplinas de otras en el proceso de \emph{reducción científica}; se trata más bien de \emph{enfocar} sobre aspectos concretos del problema global, que permiten obtener respuestas parciales, que pueden ser ensambladas más adelante de un modo holístico.

Situándonos en el plano biológico, es conocido que la enfermedad de Alzhéimer cursa con un empobrecimiento de las estructuras neuronales. Las neuronas, las células más importantes en el cerebro, tienen una estructura bien definida, con un \emph{núcleo} (o \emph{soma}) del que surgen una proyección longitudinal llamada \emph{axón} y una serie de ramificaciones, conocidas como \emph{dendritas}. La comunicación entre neuronas tiene una sola dirección y se establece siempre desde el axón de una neurona hacia las dendritas de la siguiente. Es a lo largo de la estructura de una dendrita donde encontramos los llamados \emph{contactos sinápticos}, es decir los lugares físicos en los que se establece la comunicación entre dos neuronas (en algunas zonas del cerebro, como el hipocampo, una dendrita puede recibir en promedio 1000 contactos de otras tantas neuronas). Un tipo especial de estos contactos son  unas estructuras con forma de champiñón
denominadas \emph{espinas}.  Estas estructuras están funcionalmente relacionadas con el proceso de la memoria y el aprendizaje. Pues bien, en el desarrollo de ciertas enfermedades neurodegenerativas, como es el caso del Alzhéimer, se aprecia que los contactos sinápticos disminuyen y las espinas se deterioran.

Por ello, parte de la investigación en Biología se centra en encontrar sustancias que potencien la ocurrencia de conexiones sinápticas y aumenten el número de espinas. El grupo PSE patentó un compuesto  que podría tener ese efecto de potenciador sináptico. La forma de mostrar la calidad de la patente (como paso previo a la fase de experimentación \emph{clínica}) consiste en realizar múltiples experimentos, calibrando cómo varía la densidad sináptica en presencia y en ausencia del compuesto. Para realizar esas mediciones, los biólogos toman imágenes microscópicas y en ellas, por medio de una inspección visual (aunque ayudados por tecnologías informáticas), analizan la morfología de las neuronas (que pueden ser cientos).

Es en esa fase de la investigación cuando se entra en contacto con los investigadores del grupo de Programación y Cálculo Simbólico, produciéndose la \emph{reducción} de un problema \emph{biológico} a un problema \emph{matemático}: se observa que el recuento de contactos sinápticos puede ser interpretado como un problema homológico (concretamente, de cálculo del número de componentes conexas en un espacio conveniente construido). Dicha reducción no fue un proceso fácil y lineal, sino más bien fruto de un diálogo apasionante entre biólogos y matemáticos. Intentar hacer compatibles los lenguajes empleados no fue tarea menor. Una afirmación como ``una neurona es \emph{simplemente conexa}'' no aporta mucha información a un biólogo; y, de modo simétrico, intentar dar sentido a ``los anticuerpos monoclonales son marcadores específicos'' no es cuestión evidente para un matemático.

Una vez traducido a un problema \emph{matemático}, la \emph{reducción} a un problema \emph{informático} es un proceso mucho mejor comprendido en el ámbito del Cálculo Científico: se trata de implementar algoritmos que reflejen las ideas matemáticas. Además del conocido reflujo de la informática a las matemáticas (no solo las matemáticas orientan a la informática, sino que también la implementación de los algoritmos y su ejecución profundizan en el conocimiento matemático, generando nuevas variantes y problemas, en un conocido \emph{círculo virtuoso}), en nuestro caso se tuvo la oportunidad de poder seguir interactuando con los biólogos del grupo PSE. Ellos no solo validaron experimentalmente los resultados de los programas, sino que también podían realizar nuevos experimentos para que las imágenes obtenidas en el microscopio se adecuasen mejor a las necesidades de procesamiento de los programas finales (marcando con más claridad ciertas estructuras en las imágenes, por ejemplo).

Tras esbozar brevemente al camino que va desde un problema \emph{social} hasta un problema
\emph{informático}, conviene describir en qué consiste dicho problema informático. Hablando de un modo general, se puede caracterizar como un problema de \emph{reconocimiento de patrones}: en una imagen
con ruido, se trata de reconocer ciertas estructuras (neuronas, dendritas, sinapsis, espinas, etc.).
El interés del problema reside en que, a diferencia de otras aplicaciones estándar de reconocimiento
de patrones, no se dispone de un catálogo de modelos a reconocer (piénsese en la diferencia con
los \emph{OCR}, los reconocedores de caracteres). Una neurona puede ser caracterizada en una imagen,
pero no existe un ``abecedario'' de neuronas estándar contra el que comparar. Lo que se sabe de una neurona es que tienen un núcleo (una zona con geometría aproximadamente circular, y de mayor densidad
que el resto de la imagen circundante) unida a una estructura arborescente en varios niveles (las dendritas).
Para reconocer ese tipo de patrones, parece oportuno utilizar algoritmos topo-geométricos, pues topo-geométricas son las propiedades de los caracterizan. Esa idea general es la que ha sido explorada a lo largo de los últimos años, dando lugar a desarrollos que describimos en el resto del artículo.

La organización del artículo es la siguiente. Tras enumerar los problemas tratados en el apartado segundo, detallamos los conceptos matemáticos utilizados en el tercer apartado.  En el apartado cuarto abordamos el papel de la Topología Algebraica, lo que permite en el siguiente apartado presentar los desarrollos tecnológicos realizados.  Los apartados 6 y 7 están dedicados respectivamente a la validación experimental de los resultados de los programas, y a aspectos prácticos que escapan a la rigidez de la descripción matemática. El artículo termina con unas concisas conclusiones y la bibliografía.

\section{Problemas tratados}
\label{sec:problems}

Como acabamos de comentar, el objetivo final de la investigación es el tratamiento de las enfermedades neurodegenerativas y el estudio de la efectividad de distintos fármacos. Para ello se debe llevar a cabo un proceso de determinación de  diferentes elementos estructurales de las neuronas presentes en una imagen obtenida por el microscopio.  Hasta ahora esta tarea se realizaba de forma manual; se trata de un trabajo lento y tedioso, especialmente para muestras con alta densidad de elementos estructurales. Además, todo proceso manual incluye inevitablemente un grado de subjetividad, mientras que si el tratamiento fuera automático el proceso sería totalmente objetivo. Por estos motivos se planteó como objetivo principal de nuestro trabajo el desarrollo de un software que permitiera automatizar, tanto como fuera posible, los procesos llevados a cabo por los científicos experimentales. Se persigue construir un paquete completo, que permita tomar automáticamente distintas imágenes con el microscopio, realizar un primer estudio de las mismas para localizar las zonas \emph{importantes} (en las que aparecen los elementos que se deben estudiar, es decir, las neuronas), capturar de forma automática nuevas imágenes de las zonas destacables a mayor resolución y finalmente realizar un estudio más detallado de los distintos elementos estructurales (espinas y sinapsis). Con el desarrollo de este software se persigue la posibilidad de interesar a la industria farmacéutica, puesto que la automatización permite dar el paso a un procesamiento masivo de muestras (\emph{High-throughput screening}).

\begin{figure}
 \centering
 \includegraphics[width=0.7\textwidth]{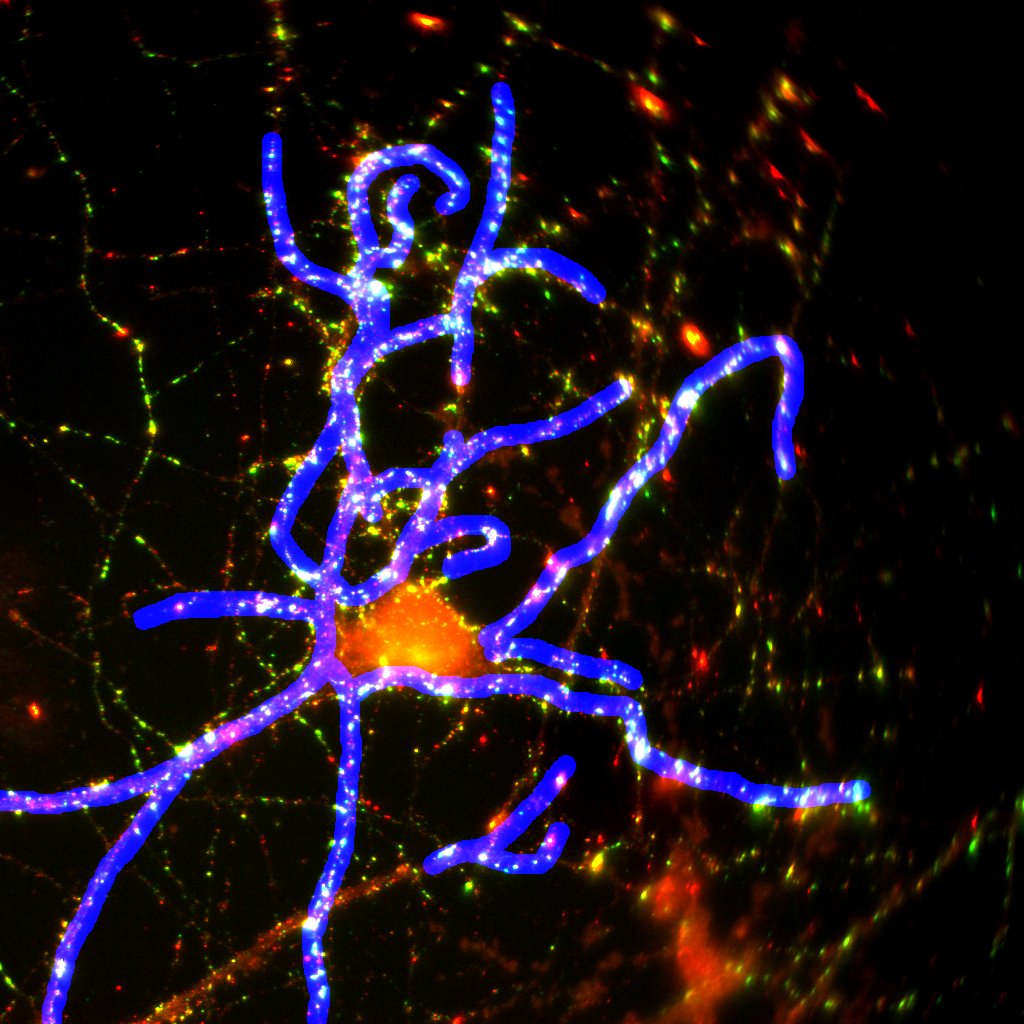}
  \caption{Recuento de sinapsis}
  \label{fig:sinapsis}
\end{figure}

Para la automatización de los procesos llevados a cabo por el grupo PSE,  se planteó en primer lugar el problema del cálculo del número de sinapsis, que son los puntos de conexión entre las neuronas. Las sinapsis tienen una gran importancia en el campo de investigación tratado ya que están relacionadas con las capacidades de cálculo del cerebro, y la posibilidad de incrementar el número de contactos sinápticos puede ser una ventaja importante en el tratamiento de enfermedades neurodegenerativas. Para realizar el recuento de sinapsis se necesitan dos imágenes diferentes de la misma neurona que se obtienen usando técnicas de laboratorio, más concretamente utilizando dos marcadores que reconocen estructuras sinápticas; esas dos imágenes se superponen y se realiza el recuento de las sinapsis. En la figura \ref{fig:sinapsis}, de tamaño $1024 \times 1024$ píxeles, o equivalentemente $228 \times 228$ micras, las sinapsis son los puntos que aparecen en blanco.

\begin{figure}
 \centering
 \includegraphics[width=0.8\textwidth]{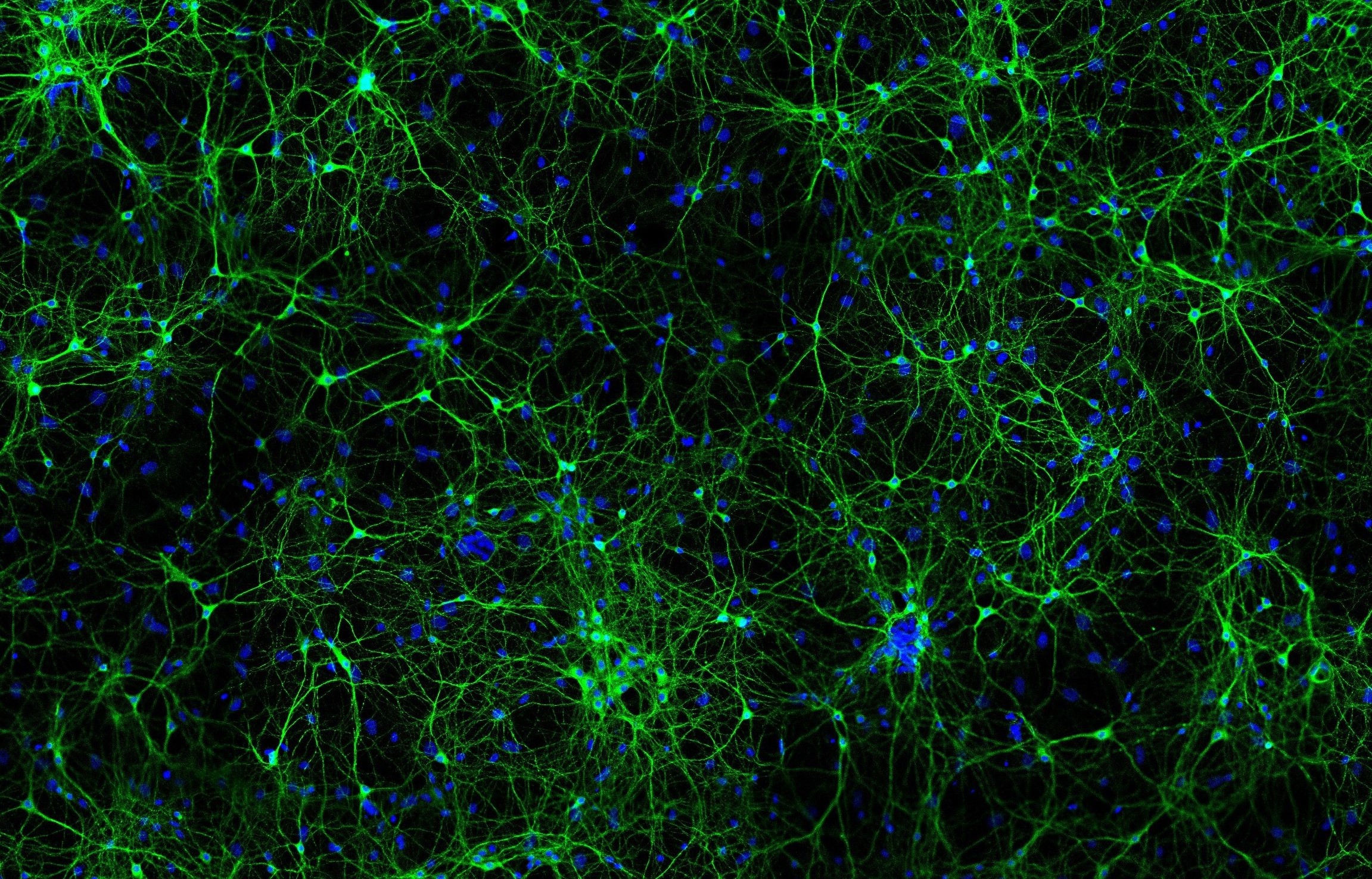}
 \caption{Estudio de la muerte neuronal}
 \label{fig:muerte_neuronal}
\end{figure}

Otro problema considerado consiste en contar el número total de neuronas que aparecen en una imagen de cultivo (véase la figura \ref{fig:muerte_neuronal}). Para ello hay que localizar los núcleos de neuronas que aparecen en la imagen y diferenciarlos de otros elementos (como \emph{astrocitos}, células que no son neuronas). Este recuento es necesario para el estudio de la muerte neuronal que se produce, por ejemplo, al sufrir un ictus. En este estudio se compara el número de núcleos disponibles antes y después de la aplicación de una sustancia a la muestra. Las imágenes consideradas en este problema tienen un tamaño de $16193 \times 10279$ píxeles, lo que en este caso equivale a $3676 \times 2333$ micras, y se obtienen como un mosaico de $8 \times 5$
imágenes de tamaño $2048 \times 2048$ píxeles (con un solapamiento del $2\%$).

\begin{figure}
 \centering
 \includegraphics[width=0.7\textwidth]{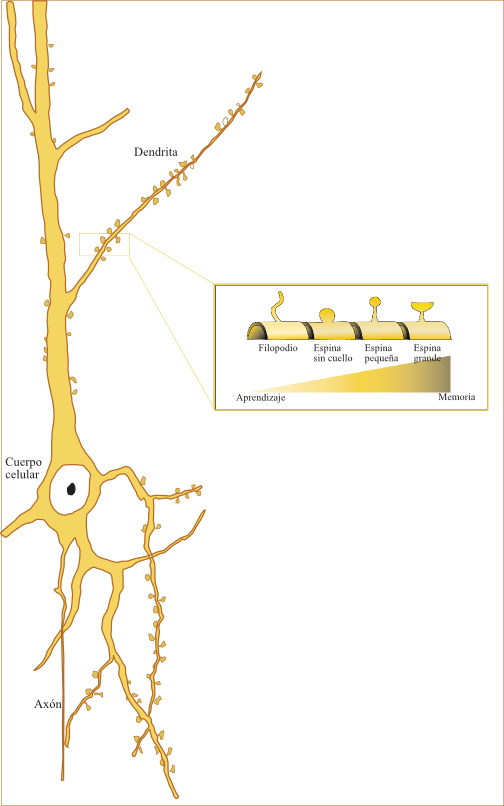}
  \caption{Clasificación morfol\'ogica de las espinas}
  \label{fig:clasificacion}
\end{figure}

Otro de los problemas a tener en cuenta en el desarrollo de medicamentos contra el Alzhéimer es el recuento y la clasificación de las espinas de una neurona. La figura~\ref{fig:clasificacion} representa (un fragmento de) una neurona; las espinas son las distintas protuberancias que sobresalen en cada dendrita. Las espinas no son estructuras rígidas, ya que pueden cambiar de forma y de tamaño de manera dinámica. El software planteado debería contar cuántas espinas aparecen y estudiar su forma, clasificando todas ellas en distintos grupos. En la figura \ref{fig:clasificacion} se representa la clasificación morfológica general de los tipos de espinas. Las espinas con cabeza pequeña están principalmente relacionadas con procesos de aprendizaje, mientras las que poseen cabezas grandes corresponden a funciones de memoria.

\begin{figure}
 \centering
 \includegraphics[width=0.8\textwidth]{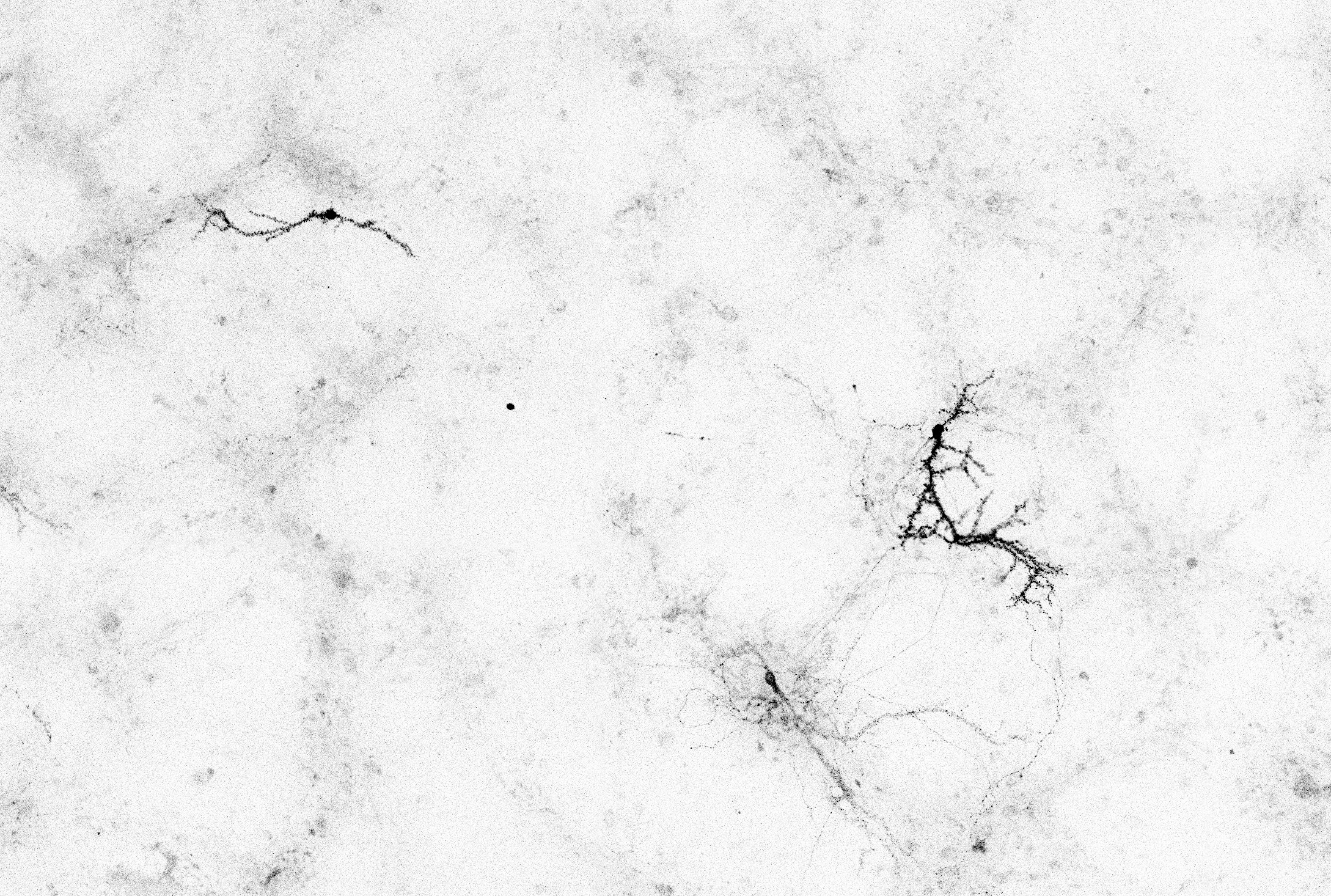}
 \caption{Imagen con varias neuronas}
 \label{fig:mosaico}
\end{figure}

Como un problema intermedio para el recuento y clasificación de espinas, se ha considerado también el problema de localizar las distintas neuronas que aparecen dentro de una imagen de cultivo. Las imágenes a estudiar en este problema tienen tamaño  $12107 \times 9174$ píxeles o $9172 \times 6950$ micras y se obtienen como un mosaico de $12\times 9$ imágenes de $1024 \times 1024$ píxeles. Este paso se utilizará como pieza necesaria para el software de recuento y clasificación de espinas; una vez que se tiene la localización de cada neurona, se debería captar otra imagen con una resolución mayor sobre las coordenadas correspondientes y a partir de la nueva foto realizar el cálculo y la clasificación de espinas deseada. En la figura \ref{fig:mosaico}, que representa un pequeño fragmento de las imágenes tratadas, pueden observarse tres neuronas.

\begin{figure}
 \centering
 \begin{tikzpicture}
\draw (0,0) node{\includegraphics[scale=.07]{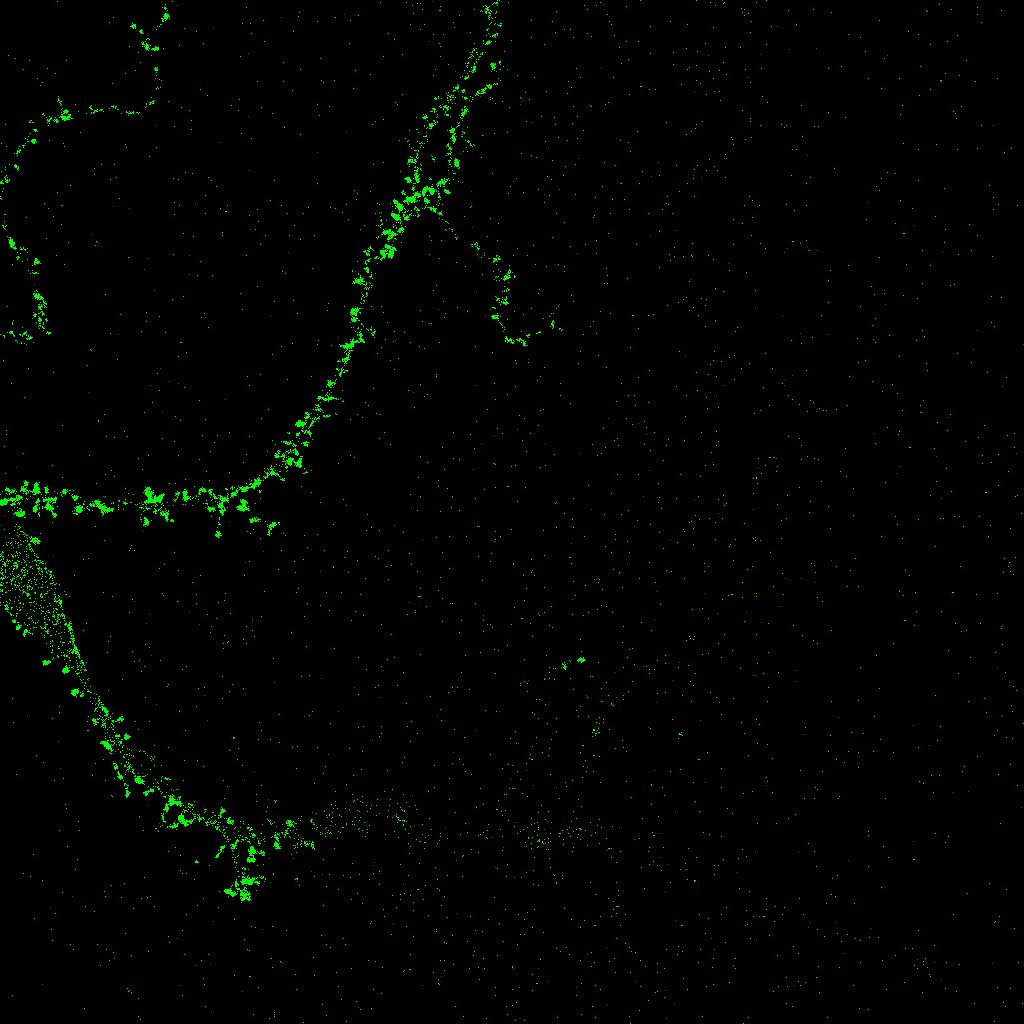}};
\draw (2.7,0) node{\includegraphics[scale=.07]{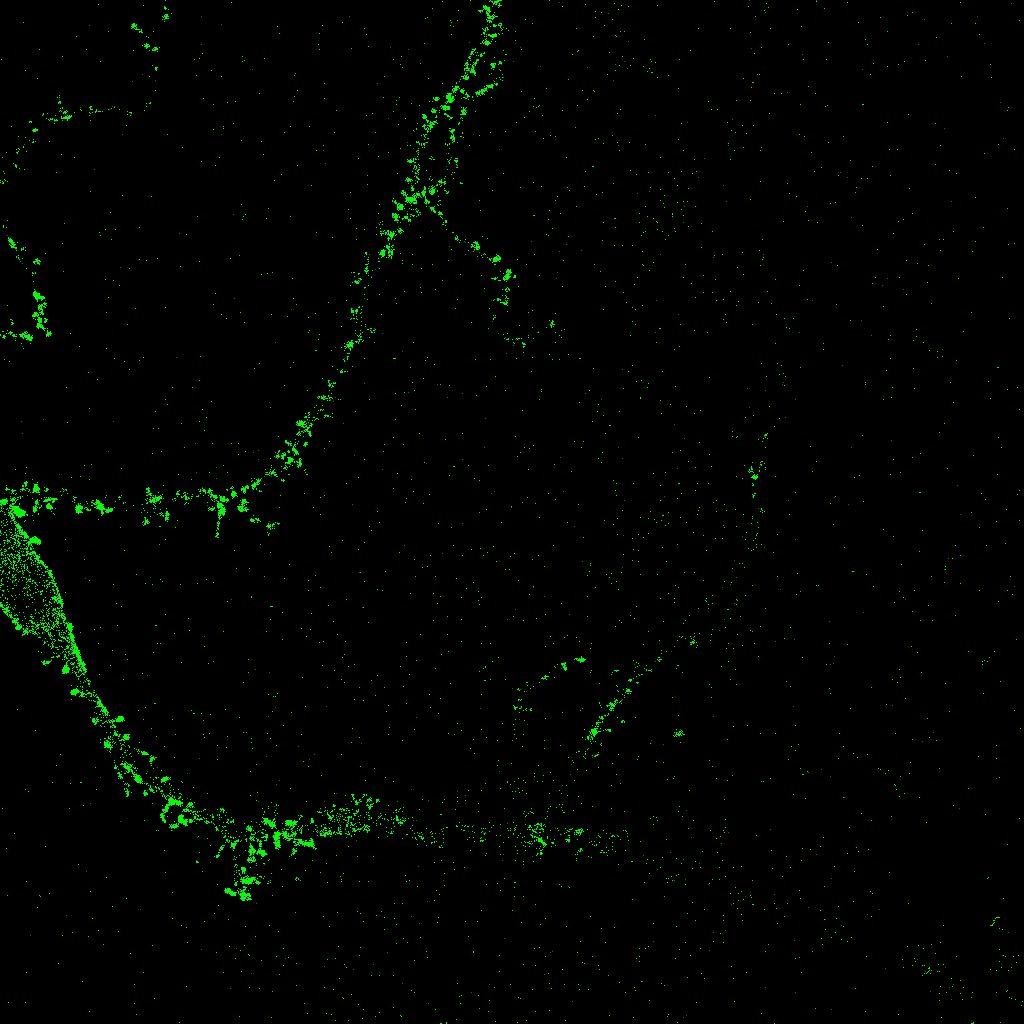}};
\draw (5.4,0) node{\includegraphics[scale=.07]{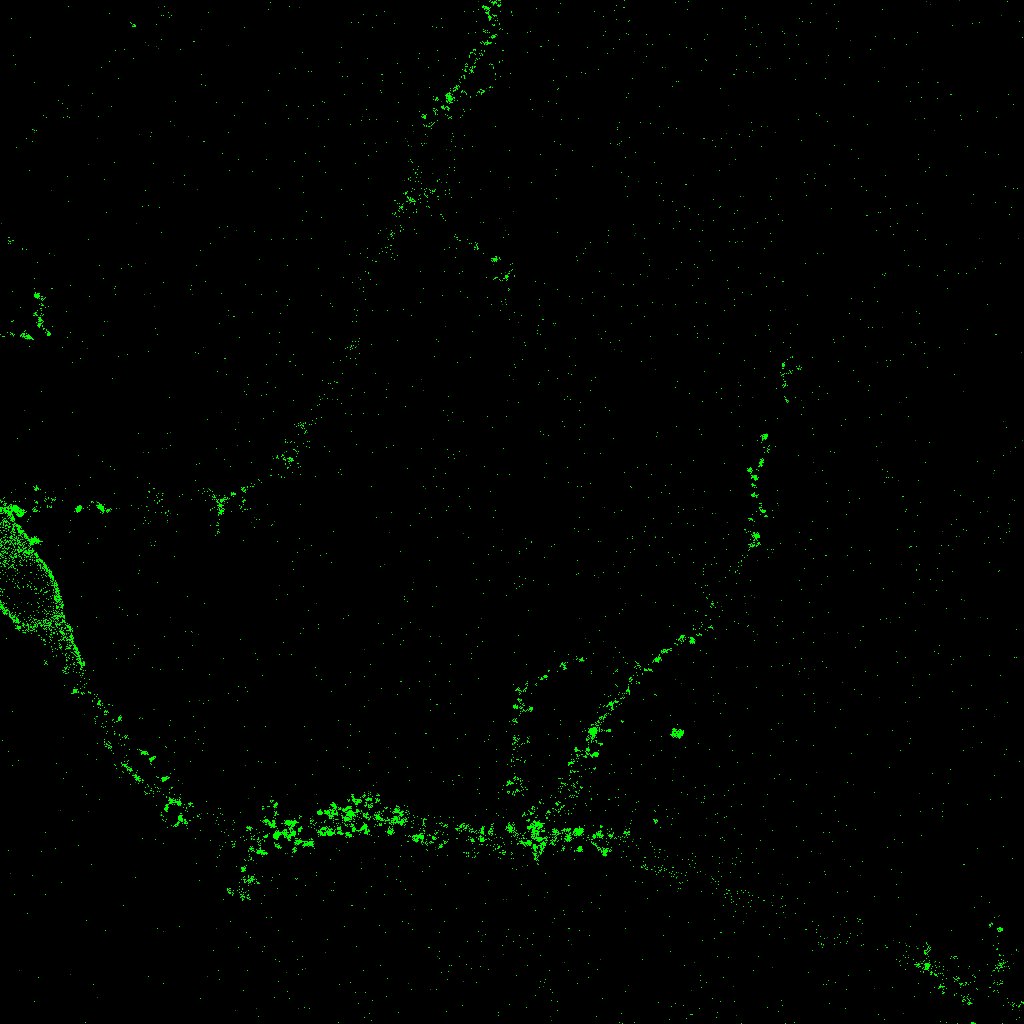}};
\draw (8.1,0) node{\includegraphics[scale=.07]{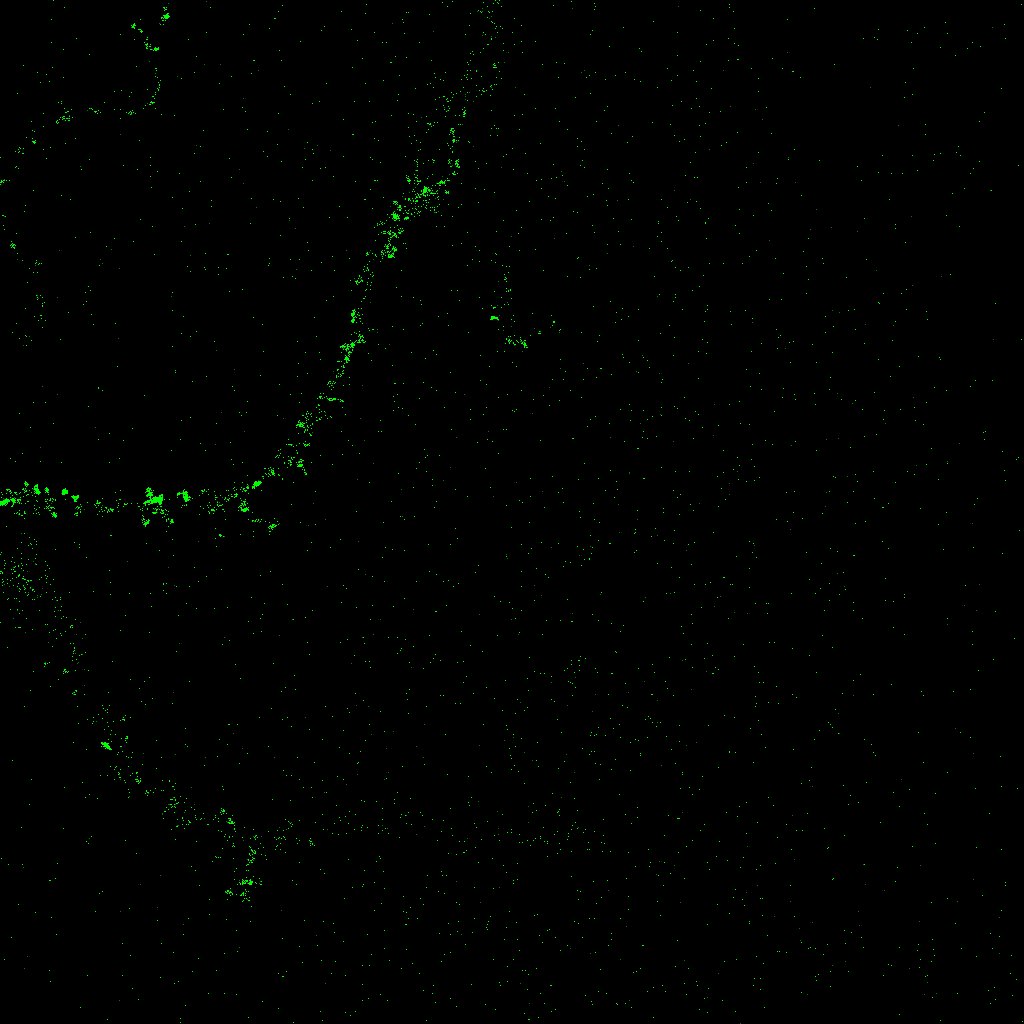}};
\draw (0,2.7) node{\includegraphics[scale=.07]{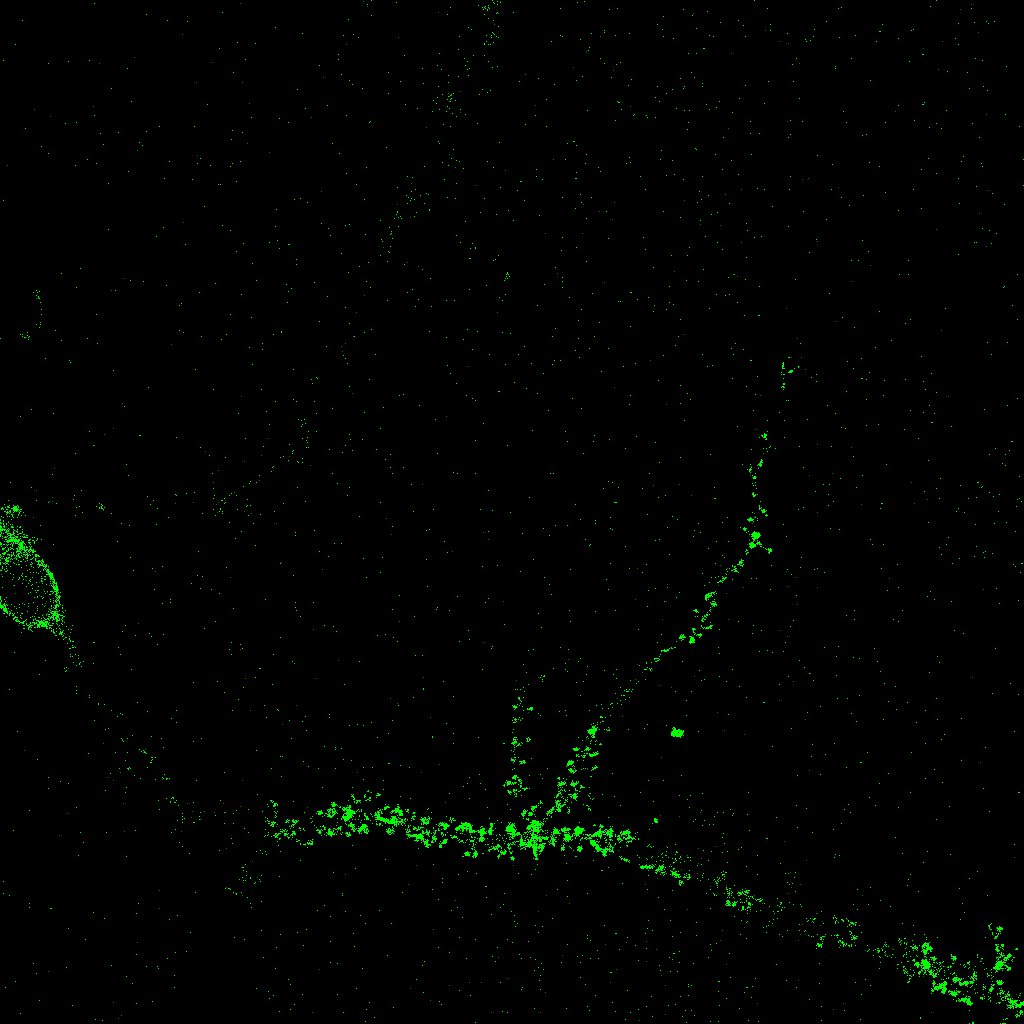}};
\draw (2.7,2.7) node{\includegraphics[scale=.07]{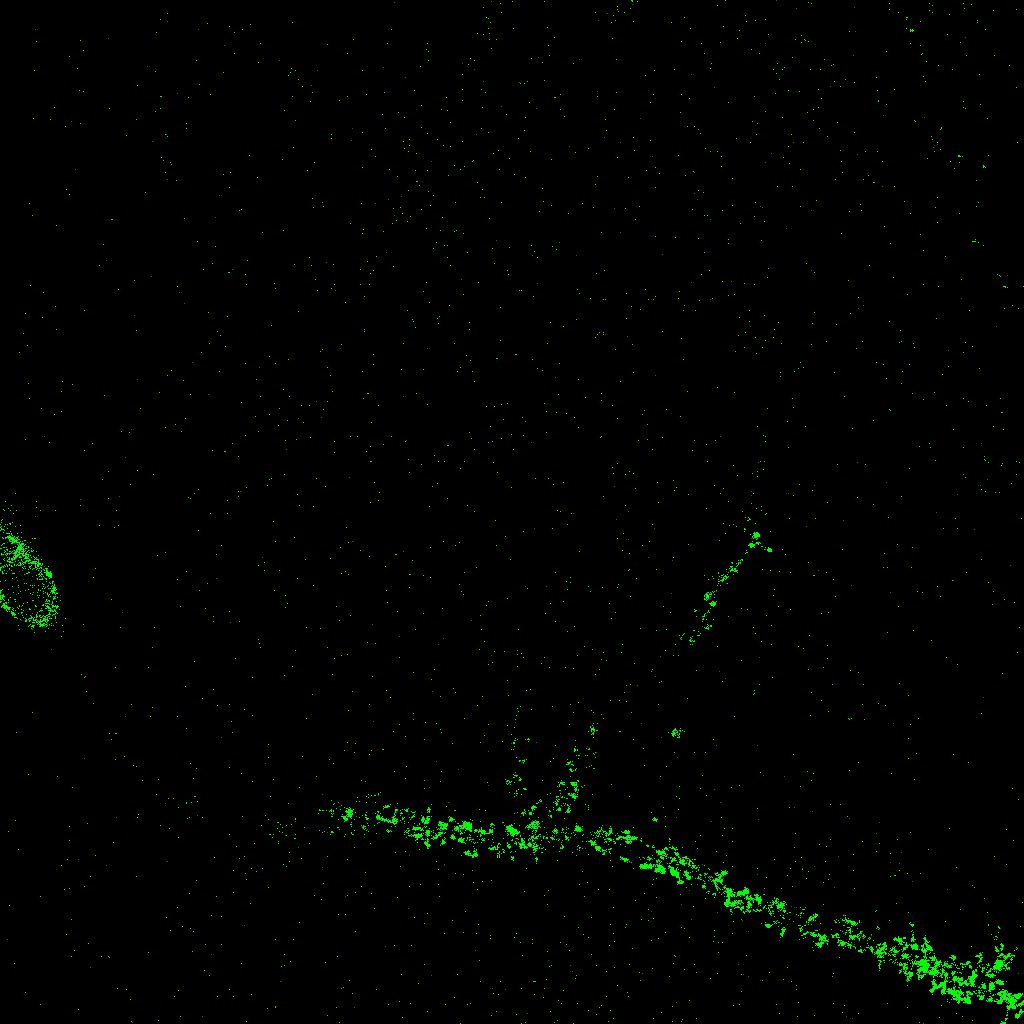}};
\draw (5.4,2.7) node{\includegraphics[scale=.07]{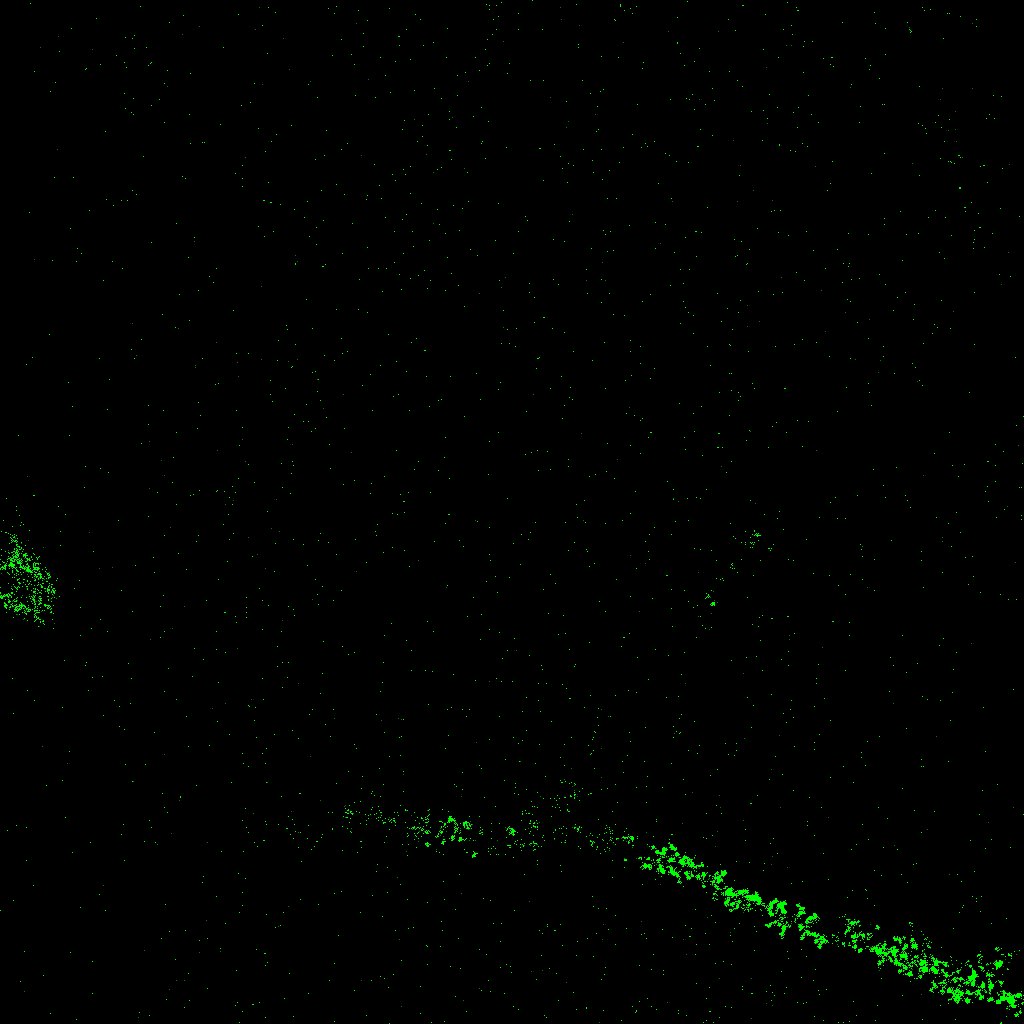}};
\draw (8.1,2.7) node{\includegraphics[scale=.07]{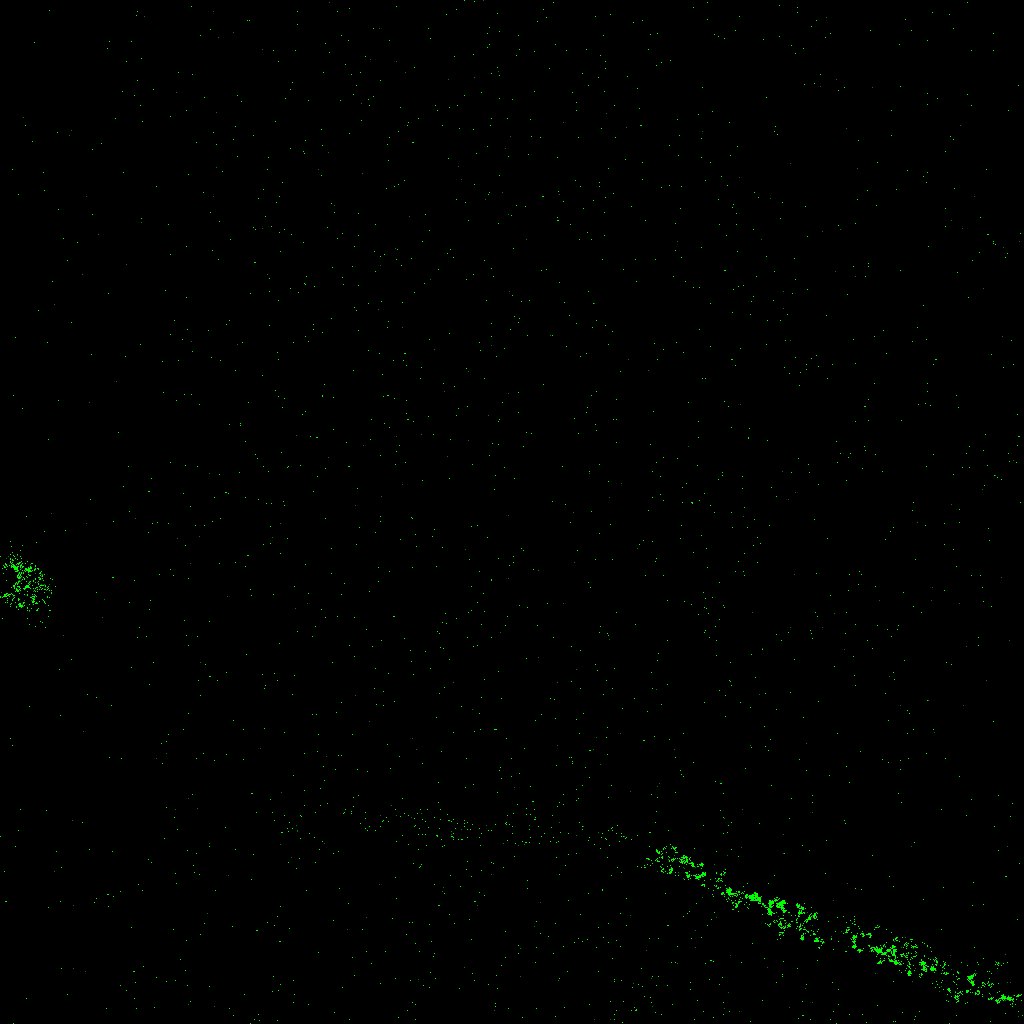}};
\end{tikzpicture}
 \caption{Pila de imágenes en distintos niveles}
 \label{fig:stack}
\end{figure}

Por último, se ha planteado también el problema de la localización de la estructura de una neurona a partir de una pila de imágenes. Partiendo de una serie de imágenes 2D correspondientes a distintas alturas (figura \ref{fig:stack}, en este caso de tamaño $1024 \times 1024$ píxeles o $49 \times 49$ micras), se trata de realizar la reconstrucción 3D determinando la estructura de la neurona y descartando otros elementos \emph{irrelevantes}. 

\section{Conceptos matemáticos utilizados}
\label{sec:conceptos_mates}

Tratando de cubrir las necesidades del grupo de biólogos PSE, se ha considerado el desarrollo de diversos programas que resuelvan los problemas planteados en la sección \ref{sec:problems}. Además de utilizar técnicas estándar en procesamiento de imágenes digitales (filtrados, eliminación de ruido, etc.) basados en distintos conceptos estadísticos como la moda, la media y la mediana,
se han utilizado diferentes ideas topo-geométricas. Presentamos en esta sección los aspectos matemáticos específicos de nuestros problemas.

Uno de los conceptos  más importantes para nuestro trabajo es la noción de \emph{conexión}, o más concretamente, el cálculo del número de componentes conexas. Ésta es la idea fundamental empleada para el recuento de sinapsis en una imagen de cultivo: tras solapar dos imágenes diferentes de la misma neurona en un mismo instante obtenidas tras la aplicación de dos marcadores primarios y señalar manualmente la región en la que debe realizarse la medida, las distintas sinapsis corresponden a las componentes conexas obtenidas en la intersección. El programa realiza este cálculo, marca sobre la imagen las distintas apariciones y nos da el número total de sinapsis (componentes conexas) localizadas. 

La idea de conexión aparece también en el problema del cálculo del número de núcleos (o neuronas) en imágenes de inmunoflorescencia. En este caso, se parte de una imagen en dos canales (correspondientes a núcleos celulares y neuronas), y a partir de las componentes conexas obtenidas en la intersección se aplican diversos criterios geométricos  para seleccionar las que corresponden a núcleos de neuronas (y desechar, por ejemplo, las que corresponden a n\'ucleos de astrocitos u otro tipo de células del sistema nervioso que no interesa considerar). Una vez realizada la selección, el número total de componentes conexas nos determina el número de neuronas disponibles en la imagen estudiada. Entre los criterios geométricos utilizados para la selección, se considera  la forma de las componentes conexas (eliminando las que son \emph{oblongas}, esto es, que la diferencia de tamaño entre el eje mayor y el eje menor sea mayor que 2) y el tamaño (se desechan las de más 200 píxeles, que suelen corresponder a astrocitos o \emph{cl\'usteres} de n\'ucleos, y las de menos de 40 píxeles, que se consideran \emph{ruido}). También se eliminan aquellas componentes que se encuentran en una zona muy \emph{densa}. Para ello se realizan diferentes círculos concéntricos alrededor de la misma comprobando si la moda va disminuyendo; si no decrece, el posible núcleo se desecha por corresponder a una \emph{nube} de n\'ucleos (o tambi\'en un cl\'uster).

Subyacente a la idea de \emph{conexión} empleada hasta ahora se encuentra el concepto de \emph{continuidad} (que en imágenes digitales es muy próximo al de \emph{contig\"uidad}). Aunque las neuronas estudiadas para el tratamiento de las enfermedades neurológicas son una estructura continua, las fotos obtenidas por el microscopio no lo son; aparecen una gran cantidad de puntos dispersos. El ojo humano hace continua esta estructura de manera natural, y es capaz de observar rápidamente la disposición de las diferentes neuronas sobre la imagen, ``uniendo'' los puntos que aparecen próximos. Sin embargo, el ordenador no es capaz de hacerlo automáticamente. Para el desarrollo de los programas para el recuento del número de sinapsis y el número de núcleos es necesario aplicar un preprocesamiento previo mediante diferentes filtros que hacen (más) continua la imagen. Para la localización de neuronas, utilizamos además una técnica de seguimiento de caminos permitiendo \emph{saltos} de dos o tres píxeles. Emerge así también la noción de \emph{distancia} en un espacio discreto, y la de \emph{adyacencia generalizada} (en la que dos píxeles son declarados contiguos si est\'an suficientemente pr\'oximos).

En el problema de la localización de neuronas aparece también el concepto de árbol. A partir de ciertos puntos iniciales (los de mayor intensidad), se traza un cuadrado de tamaño $25\times25$ píxeles alrededor de él y se van considerando los cuadrados adyacentes al anterior. Se seleccionan aquellos en los que se localiza un camino de longitud mayor o igual que $15$ píxeles y se sigue buscando en los cuadrados adyacentes a los seleccionados. Para decidir el orden de exploración de los diferentes cuadrados se sigue un proceso de búsqueda en anchura, mientras que para localizar los caminos dentro de cada cuadrado se sigue un proceso de búsqueda en profundidad. La determinación de los valores $(25\times 25, 15)$ es realizada de modo empírico, guiado por el conocimiento experto de los biólogos.

\section{El papel de la Topología Algebraica}
\label{sec:algoritmos}

Para la implementación de las ideas explicadas en la sección \ref{sec:conceptos_mates} se han utilizado diversos algoritmos y herramientas de Topología Algebraica. Aunque en ocasiones hay otros métodos para realizar esos cálculos que pueden expresarse en términos más elementales, el enfoque homológico fue el privilegiado en el proyecto europeo ForMath \cite{Formath}, cuyo objetivo era desarrollar programas \emph{con correcci\'on verificada}, pues as\'{\i} se reutiliz\'o trabajo previo en torno a la verificación de algoritmos en Topología Algebraica. Ese punto de vista se ha visto posteriormente reforzado al aplicar conceptos como el de \emph{homología persistente} en nuestro ámbito.

\begin{figure}
 \centering
\begin{tikzpicture}[framed,scale=.4] 
\shorthandoff{>}

\draw (-.5,3) node[color=black]{Imagen digital};
\draw (-.5,-11.5) node[color=black]{Complejo simplicial};
\draw (14,-11.5) node[color=black]{Complejo de cadenas};
\draw (14,3) node[color=black]{Grupos de homología};

\draw[fill=white,draw=none] (-3,-3) rectangle (2,2);
\draw[step=1,gray!20!white,very thin,fill=white] (-3,-3) grid (2,2);
\draw[fill=black,draw=none] (-2,2) rectangle (-1,1);
\draw[fill=black,draw=none] (1,2) rectangle (2,1);
\draw[fill=black,draw=none] (-3,1) rectangle (-2,0);
\draw[fill=black,draw=none] (-1,1) rectangle (0,0);
\draw[fill=black,draw=none] (-2,0) rectangle (-1,-1);
\draw[fill=black,draw=none] (0,0) rectangle (1,-1);
\draw[fill=black,draw=none] (-1,-1) rectangle (0,-2);
\draw[fill=black,draw=none] (1,-1) rectangle (2,-2);
\draw[fill=black,draw=none] (0,-2) rectangle (1,-3);

\draw[->] (0,-3.5) -- (0,-5.5);

\draw[color=black,fill=black!20!white] (-2,-6) rectangle (-1,-7);
\draw[color=black,fill=black!20!white] (1,-6) rectangle (2,-7);
\draw[color=black,fill=black!20!white] (-3,-7) rectangle (-2,-8);
\draw[color=black,fill=black!20!white] (-1,-7) rectangle (0,-8);
\draw[color=black,fill=black!20!white] (-2,-8) rectangle (-1,-9);
\draw[color=black,fill=black!20!white] (0,-8) rectangle (1,-9);
\draw[color=black,fill=black!20!white] (-1,-9) rectangle (0,-10);
\draw[color=black,fill=black!20!white] (1,-9) rectangle (2,-10);
\draw[color=black,fill=black!20!white] (0,-10) rectangle (1,-11);

\draw[color=black] (-2,-6) -- (-1,-7);
\draw[color=black] (1,-6) -- (2,-7);
\draw[color=black] (-3,-7) -- (-2,-8);
\draw[color=black] (-1,-7) -- (0,-8);
\draw[color=black] (-2,-8) -- (-1,-9);
\draw[color=black] (0,-8) -- (1,-9);
\draw[color=black] (-1,-9) -- (0,-10);
\draw[color=black] (1,-9) -- (2,-10);
\draw[color=black] (0,-10) -- (1,-11);

\draw[color=black,fill=black] (-2,-6) circle (2pt and 2pt);
\draw[color=black,fill=black] (1,-6) circle (2pt and 2pt);
\draw[color=black,fill=black] (-3,-7) circle (2pt and 2pt);
\draw[color=black,fill=black] (-1,-7) circle (2pt and 2pt);
\draw[color=black,fill=black] (-2,-8) circle (2pt and 2pt);
\draw[color=black,fill=black] (0,-8) circle (2pt and 2pt);
\draw[color=black,fill=black] (-1,-9) circle (2pt and 2pt);
\draw[color=black,fill=black] (1,-9) circle (2pt and 2pt);
\draw[color=black,fill=black] (0,-10) circle (2pt and 2pt);

\draw[color=black,fill=black] (-1,-6) circle (2pt and 2pt);
\draw[color=black,fill=black] (2,-6) circle (2pt and 2pt);
\draw[color=black,fill=black] (-2,-7) circle (2pt and 2pt);
\draw[color=black,fill=black] (0,-7) circle (2pt and 2pt);
\draw[color=black,fill=black] (-1,-8) circle (2pt and 2pt);
\draw[color=black,fill=black] (1,-8) circle (2pt and 2pt);
\draw[color=black,fill=black] (0,-9) circle (2pt and 2pt);
\draw[color=black,fill=black] (2,-9) circle (2pt and 2pt);
\draw[color=black,fill=black] (1,-10) circle (2pt and 2pt);

\draw[color=black,fill=black] (-1,-7) circle (2pt and 2pt);
\draw[color=black,fill=black] (2,-7) circle (2pt and 2pt);
\draw[color=black,fill=black] (-2,-8) circle (2pt and 2pt);
\draw[color=black,fill=black] (0,-8) circle (2pt and 2pt);
\draw[color=black,fill=black] (-1,-9) circle (2pt and 2pt);
\draw[color=black,fill=black] (0,-10) circle (2pt and 2pt);
\draw[color=black,fill=black] (2,-10) circle (2pt and 2pt);
\draw[color=black,fill=black] (1,-11) circle (2pt and 2pt);

\draw[color=black,fill=black] (-2,-7) circle (2pt and 2pt);
\draw[color=black,fill=black] (1,-7) circle (2pt and 2pt);
\draw[color=black,fill=black] (-3,-8) circle (2pt and 2pt);
\draw[color=black,fill=black] (-1,-8) circle (2pt and 2pt);
\draw[color=black,fill=black] (-2,-9) circle (2pt and 2pt);
\draw[color=black,fill=black] (-1,-10) circle (2pt and 2pt);
\draw[color=black,fill=black] (1,-10) circle (2pt and 2pt);
\draw[color=black,fill=black] (0,-11) circle (2pt and 2pt);

 \draw[->] (3,-9) -- (9,-9);

 \draw (10,-8) node[anchor=west]{{\large $C_0=$ vértices}};
 \draw (10,-9) node[anchor=west]{{\large $C_1=$ aristas}};
 \draw (10,-10) node[anchor=west]{{\large $C_2=$ triángulos}};

 \draw[->] (14,-7) -- (14,-1.5);
 \draw (10,-.5) node[anchor=west]{{\large $H_1=\mathbb{Z}\oplus \mathbb{Z} \oplus \mathbb{Z}$}};
 \draw (10,.5) node[anchor=west]{{\large $H_0=\mathbb{Z}\oplus \mathbb{Z}$}};
  \draw[->] (9.5,0) -- (3,0);
\end{tikzpicture}
 \caption{Cálculo de componentes conexas mediante homolog\'ia}
 \label{fig:homologia}
\end{figure}
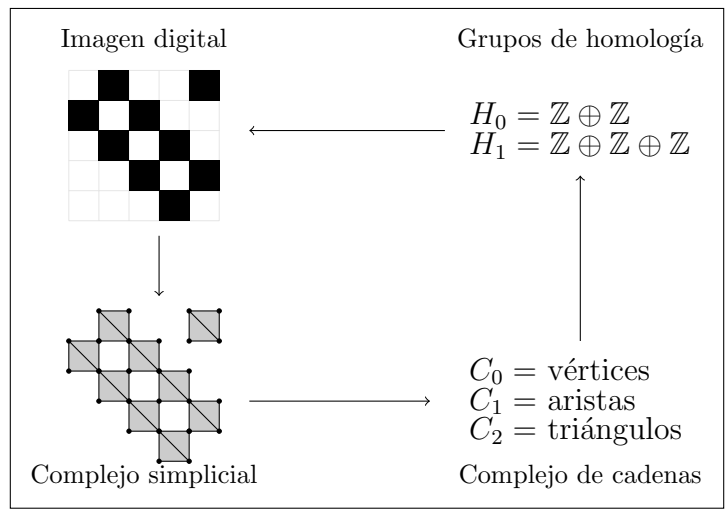

Por ejemplo, el cálculo de componentes conexas puede realizarse mediante el cálculo de grupos de homología. Dada una imagen digital, se considera el complejo celular (o simplicial) asociado. Los grupos de homología de éste en grados~$0$ y~$1$ corresponden respectivamente al número de componentes conexas y al número de \emph{agujeros} que aparecen en la imagen. Véase la figura \ref{fig:homologia}.

Para realizar el cálculo de los grupos de homología podemos \emph{reducir} el tamaño del complejo celular o simplicial haciendo uso de la noción de \emph{campo vectorial discreto}, una herramienta de Topología Diferencial discreta introducida por Robin Forman en \cite{For98} y de gran utilidad en el entorno algebraico y de la \emph{homología efectiva}~\cite{RS10}. Dado un complejo celular $C$, un campo vectorial discreto $V$ es una lista de pares de elementos de $C$ donde el primer elemento es una cara \emph{regular} del segundo y tal que cada elemento de $C$ aparece como mucho una vez en $V$.  En una imagen digital, un campo vectorial discreto corresponde a un conjunto de \emph{flechas} relacionando dos elementos, el destino de una dimensión mayor que la salida (de vértices a aristas; de aristas a píxeles).

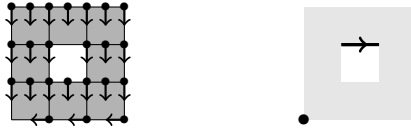
\begin{figure}
 \centering
\begin{tikzpicture}
[scale = 0.5, baseline = 0.7cm]
\shorthandoff{>}
 \fill [black!30] (0,0) rectangle (3,3) ;
 \fill [white] (1,1) rectangle (2,2) ;
 \foreach \i in {0,...,3}
   {\draw (\i, 0) -- +(0,3) ;
    \draw (0, \i) -- +(3,0) ;
   }
 \foreach \i in {0,0.5,...,3}
   \foreach \j in {1,3}
     {\draw [->, thick] (\i,\j) node {\scriptsize\(\bullet\)} -- +(0,-0.5) ;}
 \foreach \i in {0,0.5,1,2,2.5,3}
   {\draw [->, thick] (\i,2) node {\scriptsize\(\bullet\)} -- +(0,-0.5) ;}
 \foreach \i in {1,2,3}
   {\draw [->, thick] (\i,0) node {\scriptsize\(\bullet\)} -- +(-0.5,0) ;}
\end{tikzpicture}
\hspace{50pt}
\begin{tikzpicture} [scale = 0.5, baseline = 0.7cm]
\shorthandoff{>}
 \fill [black!10] (0,0) rectangle (3,3) ;
 \fill [white] (1,1) rectangle (2,2) ;
 \draw [very thick] (1,2) -- (2,2) ;
 \draw [very thick, ->] (1,2) -- (1.7,2) ;
 \node at (0,0) {\(\bullet\)} ;
\end{tikzpicture}
\caption{Campo vectorial discreto sobre una imagen $3\times 3$ y complejo crítico obtenido}
\label{fig:dvf}
\end{figure}

Consideramos por ejemplo en la figura \ref{fig:dvf} una imagen digital de tamaño $3\times 3$, y el campo vectorial dibujado sobre ella. El complejo celular asociado a la imagen inicial tiene 16 vértices, 24 aristas y 8 cuadrados. El campo vectorial definido sobre él permite reducir su tamaño, ya que los grupos de homología del complejo inicial son isomorfos a los de otro complejo más pequeño en el que sólo se incluyen las celdas \emph{críticas}, esto es, las que no aparecen en ninguno de los pares (flechas) del campo vectorial. En este caso, en el complejo crítico (imagen de la derecha en la figura \ref{fig:dvf}) hay únicamente dos celdas críticas: un vértice y una arista. Esto nos indica que la homología de la imagen inicial es isomorfa a la de una circunferencia.

Para imágenes digitales más grandes, como es el caso de nuestras fotos de neuronas, este proceso de reducción permite disminuir considerablemente el tiempo de cálculo de los grupos de homología. Utilizando heurísticas adecuadas se puede obtener de ese modo rendimientos similares a los de otros algoritmos para el cálculo de componentes conexas, basados en el recorrido de grafos (de hecho, esos algoritmos pueden ser ``le\'{\i}dos'' en el lenguaje de los campos vectoriales discretos).

Otra técnica topológica muy utilizada en el tratamiento de imágenes digitales es la noción de \emph{homología persistente} \cite{EH08}, que permite detectar las componentes \emph{importantes} de una imagen desechando las partes menos útiles que se consideran \emph{ruido}. Dada una imagen digital, se debe definir primero una filtración sobre ella. Las componentes importantes serán las que persisten en todos los niveles de la filtración; el ruido \emph{muere} en alguno de los pasos.

Las imágenes neuronales obtenidas en el microscopio confocal proporcionan una pila de imágenes 2D correspondientes a diferentes alturas. A partir de ellas, calculamos la proyección máxima y una filtración sobre ella. La homología persistente de la filtración calculada permite determinar las neuronas que aparecen en la imagen y desechar el ruido.

Como una generalización de la noción de homología persistente ha surgido también el concepto de \emph{persistencia zigzag}~\cite{CS10}. En este caso no hace falta construir una filtración por lo que podemos utilizar directamente las imágenes 2D obtenidas del microscopio. La persistencia zigzag nos da la relación entre ellas y nos permite también determinar la estructura de las neuronas presentes en la imagen estudiada.

\section{Tecnologías utilizadas}

\subsection{Fiji/ImageJ}

Siguiendo las ideas explicadas en las secciones anteriores se han desarrollado diversos programas por medio de plugins para el sistema Fiji/ImageJ. Fiji~\cite{Fiji} es un programa Java que puede ser descrito como una distribución de ImageJ \cite{ImageJ}. Estos dos programas se utilizan para procesar y analizar
imágenes biomédicas y son muy utilizados en la investigación en ciencias de la vida y biomedicina. Fiji e ImageJ son proyectos de código abierto y su funcionalidad
se puede ampliar por medio de \emph{macros} o \emph{plugins} en el lenguaje de programación Java.
Entre los plugins y macros disponibles para Fiji/ImageJ, podemos encontrar  programas para
binarizar una imagen a través de diferentes algoritmos por medio de un umbral,  homogeneizar las imágenes
a través de filtros (por ejemplo, mediante el ``filtro de la mediana''), u obtener la proyección máxima de
una pila de imágenes por diferentes criterios (el más común en este ámbito es el basado en la intensidad máxima).

Los plugins desarrollados se denominan \emph{SynapCountJ}~\cite{SynapCountJ}, \emph{NeuronPersistentJ}~\cite{NeuronPersistentJ},  \emph{NucleusJ} \cite{NucleusJ} y \emph{LocationJ}. Los tres primeros ya han sido validados por los biólogos del grupo PSE y están actualmente en fase de producción, siendo utilizados por laboratorios de diversos países europeos. El plugin LocationJ se encuentra en fase de desarrollo.

\begin{figure}
 \centering
\begin{tikzpicture}
\draw (0,0) node{\includegraphics[scale=.1]{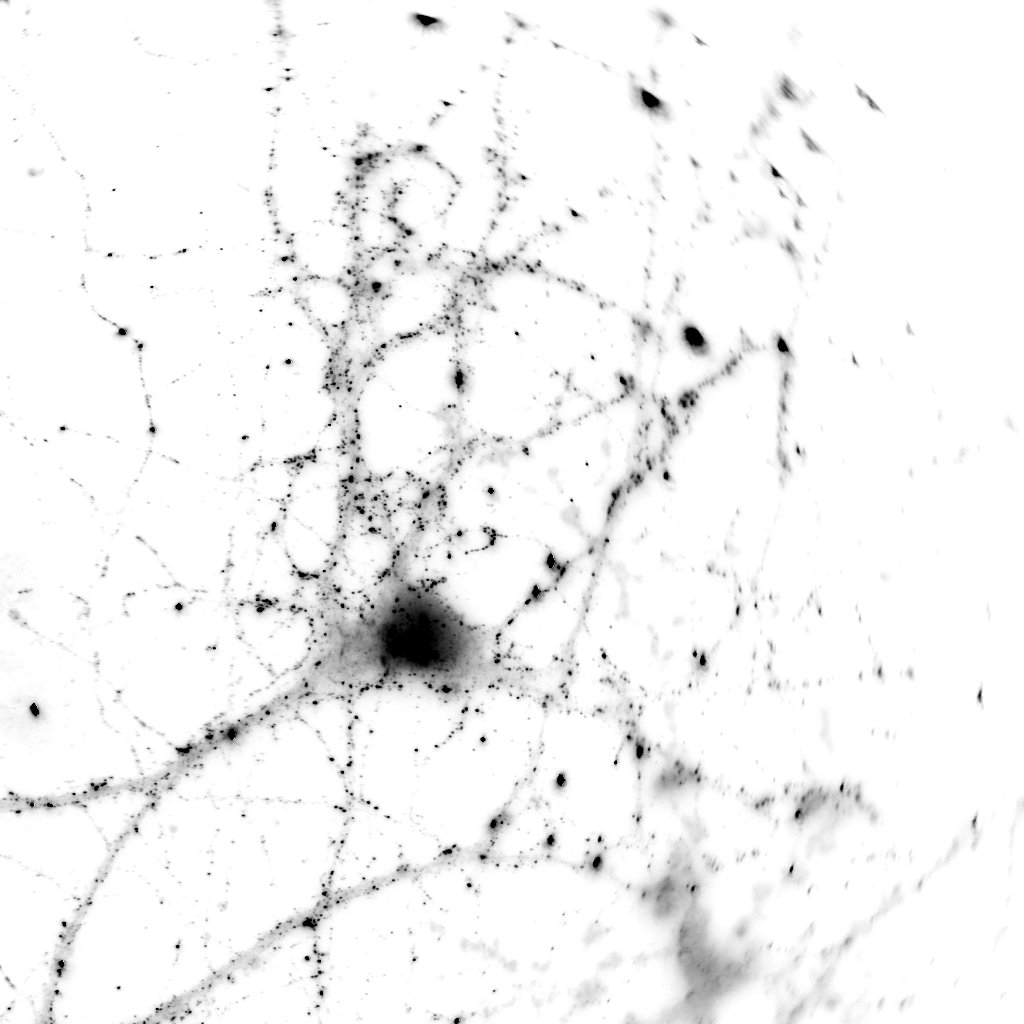}};
\draw (4,0) node{\includegraphics[scale=.1]{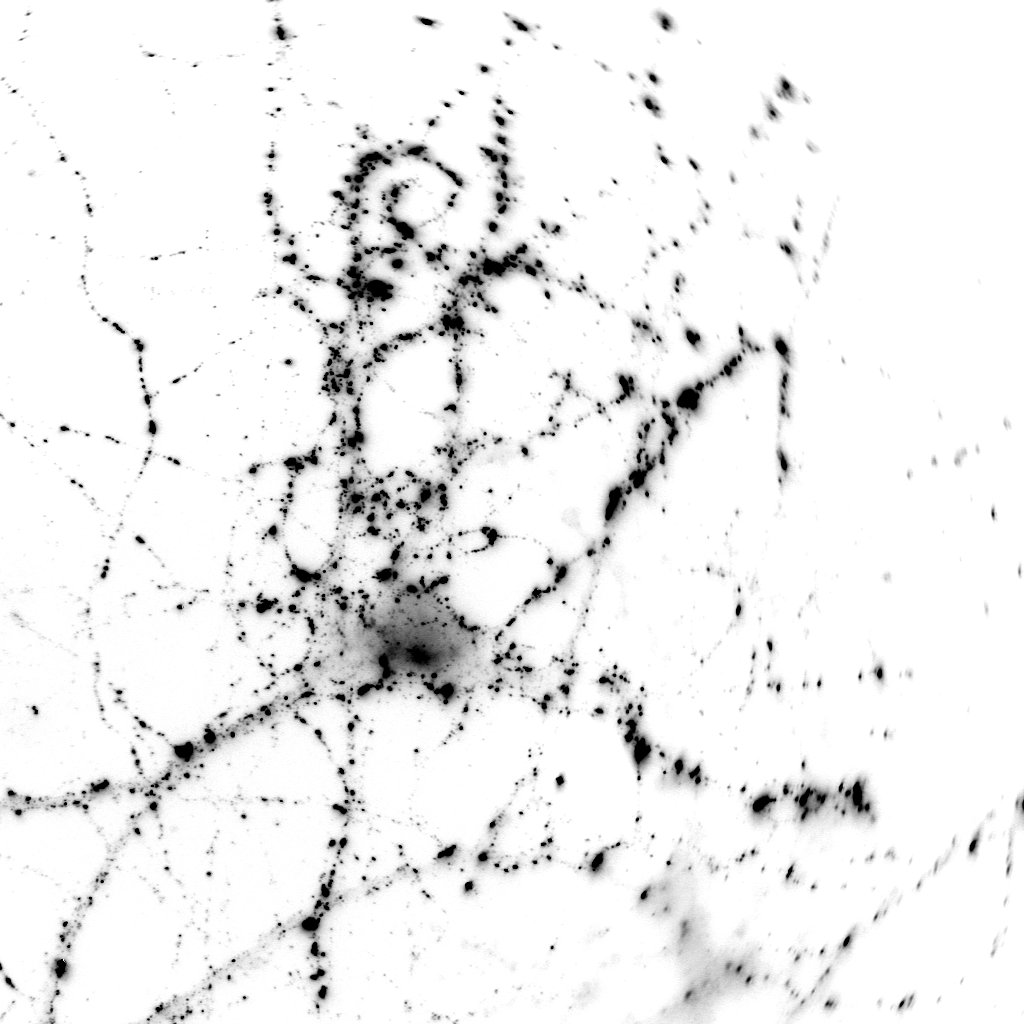}};
\draw (8,0) node{\includegraphics[scale=.1]{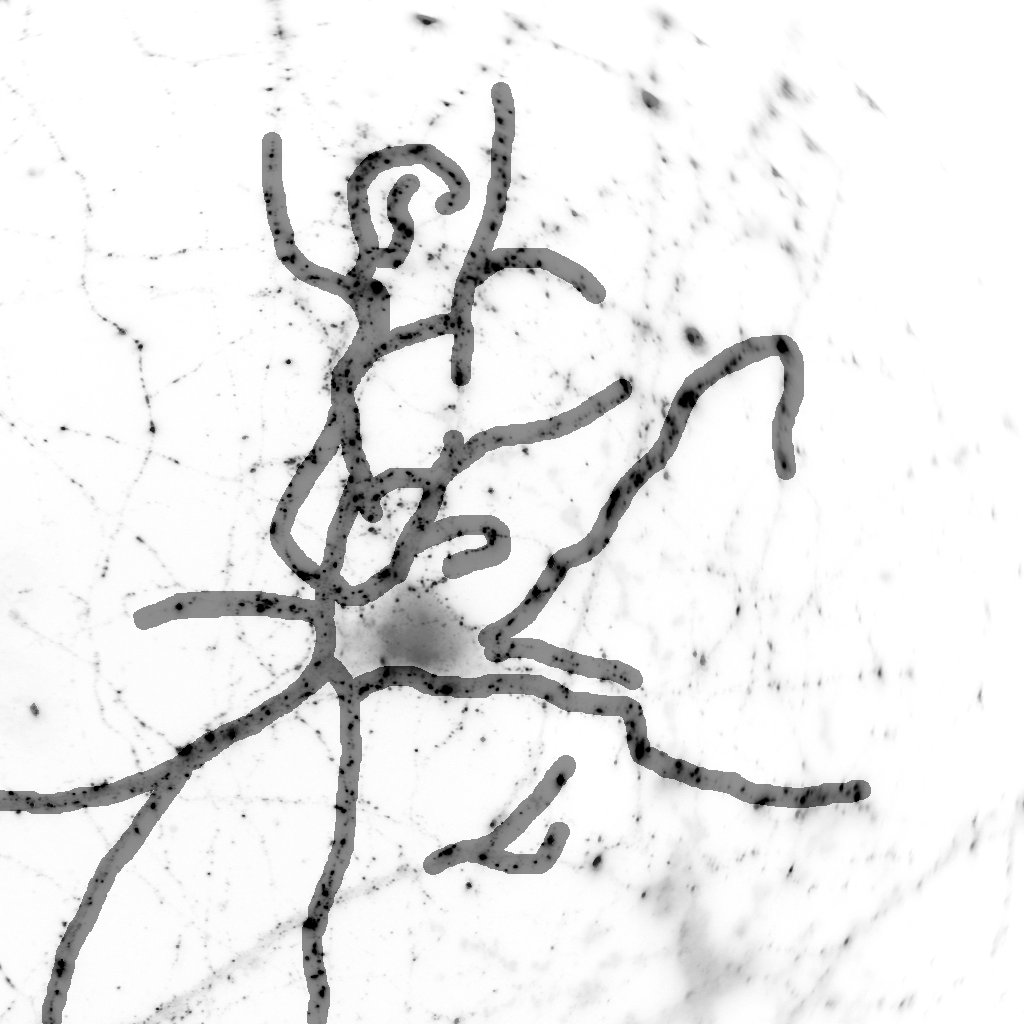}};
\end{tikzpicture}
 \caption{Resultado de SynapCountJ}
 \label{fig:synapcountj}
\end{figure}

El primer plugin que desarrollamos se denomina \emph{SynapCountJ}~\cite{SynapCountJ}, y se utiliza para calcular el número de sinapsis que aparecen en la imagen de una neurona. Como ya se ha comentado anteriormente, el programa recibe dos imágenes diferentes de la misma neurona tomadas en un momento concreto tras la aplicación de dos marcadores, en rojo y verde. El usuario debe señalar manualmente la región en la que va a realizarse la medida, esto es, las distintas ramas (dendritas) de la neurona, lo que nos da una tercera imagen en azul (para esta parte hacemos uso del plugin \emph{NeuronJ}~\cite{NeuronJ}). Con estos datos, el programa marcará como sinapsis los puntos de rojo, verde y azul que coincidan en las tres imágenes (y que aparecerán como puntos blancos, véase la figura \ref{fig:sinapsis}). El usuario puede modificar los parámetros introduciendo el rango tanto de rojo como de verde que considera que debe poseer una sinapsis. Según se va eligiendo el rango se pueden ir observando la zonas de la neurona que se van a
marcar como sinapsis para hacer una mejor estimación. Finalmente, el programa presenta la imagen con las sinapsis señaladas y muestra también una tabla en la que se indican el número de sinapsis y la densidad obtenida. Véanse las dos imágenes de partida y el resultado dado por SynapCountJ (las tres en escala de grises) en la figura \ref{fig:synapcountj}. En esta imagen las sinapsis corresponden a los puntos más oscuros dentro de la zona señalada.

\begin{figure}
 \centering
 \includegraphics[width=0.7\textwidth]{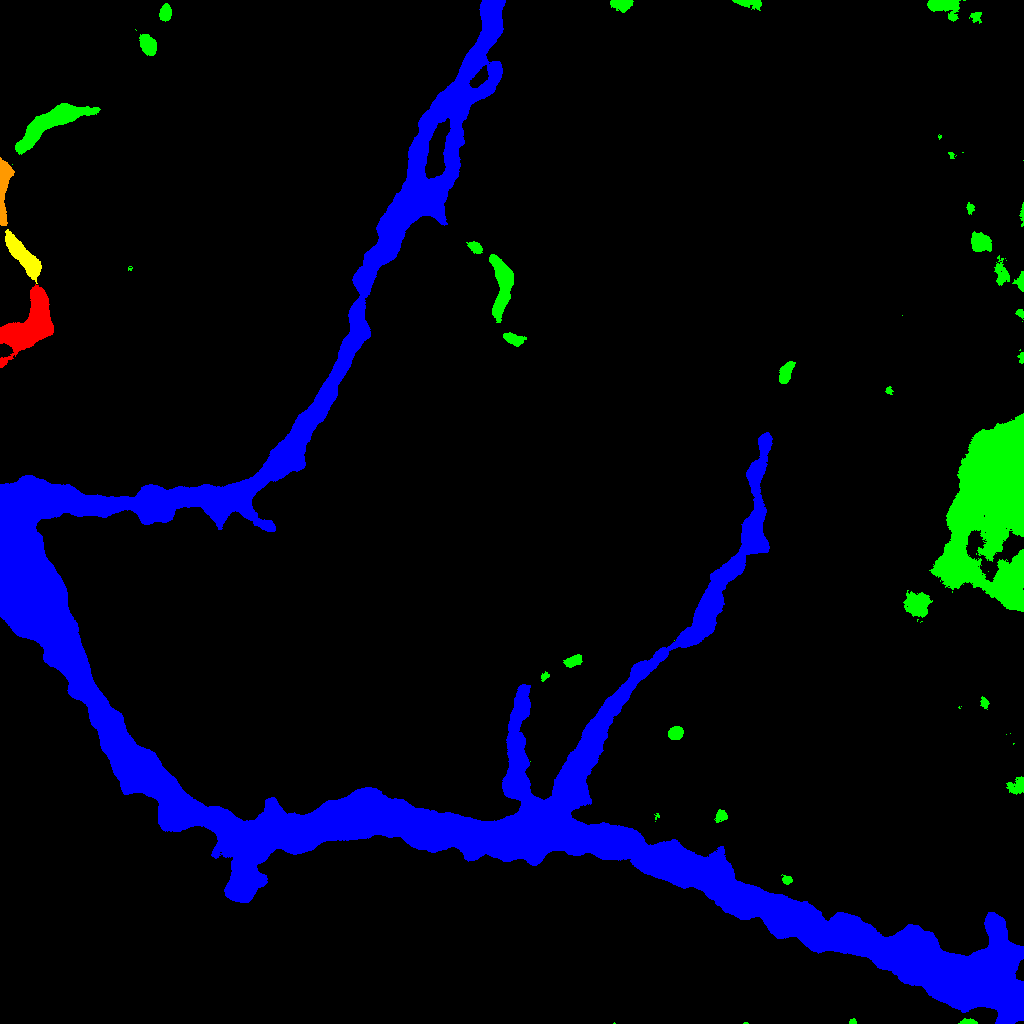}
 \caption{Resultado de NeuronPersistentJ}
 \label{fig:persistentJ}
\end{figure}

El plugin \emph{NeuronPersistentJ}~\cite{NeuronPersistentJ} se utiliza para detectar la estructura de una neurona a partir de una pila de imágenes en dos dimensiones (como en la figura~\ref{fig:stack}). En primer lugar, se procesan las imágenes con filtros que disminuyen el \emph{ruido}. En un segundo paso se aplican las ideas de homología persistente \cite{EH08} y persistencia zigzag~\cite{CS10} explicadas en la sección \ref{sec:algoritmos}. El resultado es una imagen 2D en la que se representa la estructura de la neurona, como muestra la figura \ref{fig:persistentJ}. 

\begin{figure}
 \centering
 \includegraphics[width=0.8\textwidth]{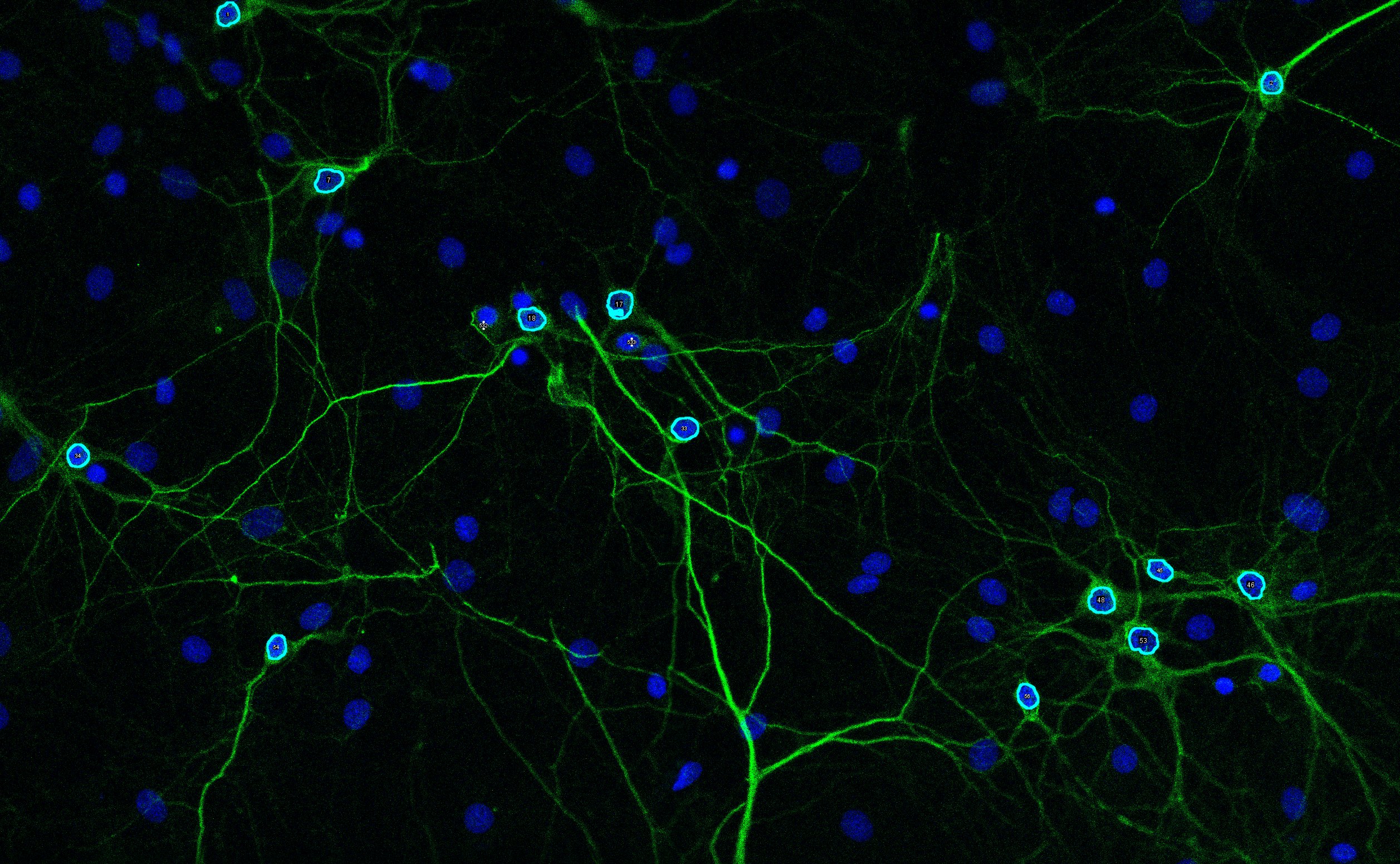}
 \caption{Resultado de NucleusJ}
 \label{fig:nucleusJ}
\end{figure}

Para el recuento y la localización de núcleos de neuronas en una imagen de cultivo se ha creado un plugin para Fiji llamado \emph{NucleusJ} \cite{NucleusJ}. En este caso, el programa parte de una imagen en dos canales que corresponden a núcleos y neuronas. En primer lugar se les aplica a ambas imágenes diferentes filtros para hacer más acertada su binarización. Tras ello se cuenta el número total de componentes conexas que aparecen en la imagen de los núcleos (número total de núcleos, ya sean de neuronas o de otras células). Después se solapan las dos imágenes y se seleccionan aquellos núcleos en los que hay intersección con la imagen de las neuronas. El programa permite especificar algunos parámetros, como el tamaño mínimo y máximo para los núcleos considerados o la diferencia máxima entre el eje mayor y el eje menor, descartando automáticamente los que no cumplen las condiciones deseadas. El plugin devuelve una imagen con los núcleos seleccionados como se muestra en la figura~\ref{fig:nucleusJ}, siendo posible modificar los parámetros y el umbral para ajustar el resultado. También se permite seleccionar manualmente más núcleos o descartar algunos de los propuestos. Finalmente, el programa informa del número total de neuronas y el número total de células obtenidos.

\begin{figure}
 \centering
 \includegraphics[width=0.8\textwidth]{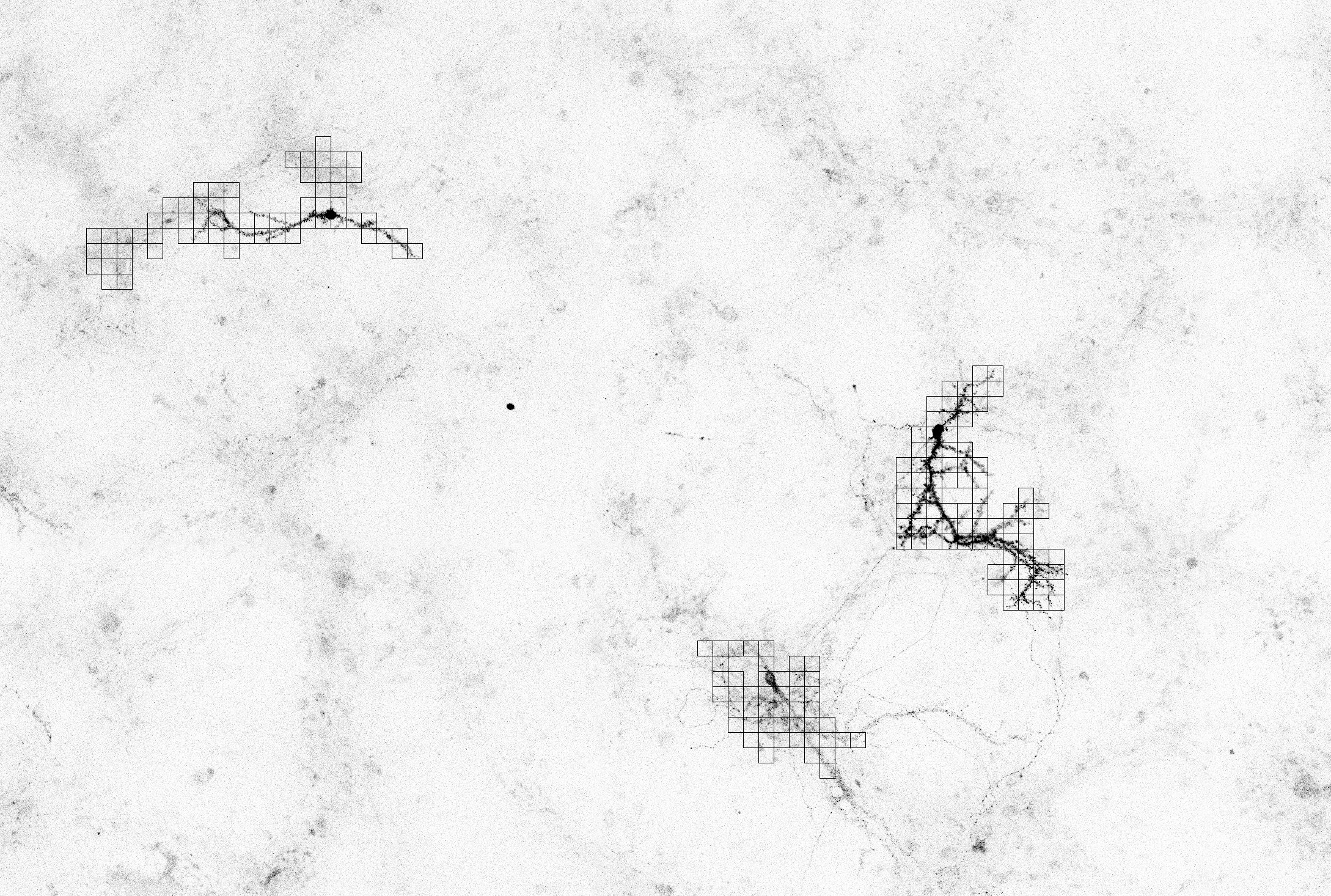}
 \caption{Resultado de LocationJ}
 \label{fig:locationJ}
\end{figure}

Por último, el plugin \emph{LocationJ}, que se encuentra actualmente en fase de desarrollo, permitirá localizar las diferentes neuronas presentes en una imagen de mosaico. Para ello se parte de los puntos de la imagen con mayor intensidad (candidatos a estar en el núcleo de alguna neurona), y se realiza una búsqueda de caminos alrededor de ellos para localizar las diferentes dendritas. Como ya se ha explicado anteriormente, para solucionar el problema de la continuidad, se permiten saltos de~2 o~3 píxeles. El programa muestra una serie de cuadrados indicando el área ocupada por la neurona, como se puede observar en la figura \ref{fig:locationJ}. Este plugin será utilizado como pieza intermedia en el programa para el recuento y clasificación de las espinas (uno de los problemas planteados por el grupo de biólogos PSE, que todavía no ha sido tratado).

Todos los nuevos plugins desarrollados para Fiji/ImageJ  son de código abierto y pueden ser obtenidos en \url{http://spineup.jimdo.com/downloads/}.

\subsection{Kenzo}

Para realizar algunos de los cálculos homológicos presentes en nuestros programas hemos utilizado la ayuda del programa \emph{Kenzo} \cite{Kenzo}. Kenzo es un sistema de cálculo simbólico en Topología Algebraica que permite el cálculo de grupos de homología y homotopía de espacios. Kenzo trabaja con estructuras algebraicas complicadas (complejos de cadenas, álgebras diferenciales graduadas, conjuntos simpliciales, grupos simpliciales, morfismos entre estos objetos, reducciones, etc.) y ha obtenido algunos resultados (por ejemplo, grupos de homología de espacios de lazos iterados de un espacio de lazos modificado por una unión de células, componentes de torres de Postnikov complejas, grupos de homotopía de espacios clasificantes suspendidos, etc.) que no eran conocidos anteriormente. Además, Kenzo ha permitido la detección de un error en un teorema publicado en \cite{MW10}, donde mediante razonamientos teóricos los autores deducen que el cuarto grupo de homotopía del espacio clasificante suspendido del grupo alternado $A_4$,  $\pi_4(\Sigma K (A_4, 1))$, es igual a $\Zset_4$; los cálculos de Kenzo han demostrado que el resultado correcto (como confirmaron posteriormente los autores de \cite{MW10}) es $\Zset_{12} $. Véase \cite{RR12}  para más detalles sobre estos cálculos.

Durante el desarrollo de los distintos plugins para Fiji/ImageJ presentados en la sección anterior se ha hecho uso de Kenzo en diversos momentos. En primer lugar, previamente a la implementación del plugin SynapCountJ, se utilizó Kenzo para comprobar que el cálculo del número de sinapsis presentes en las imágenes consideradas podía calcularse como el número de componentes conexas (y a su vez calculado a través de un grupo de homología). Después se ha utilizado este sistema de cálculo simbólico para validar los resultados obtenidos por los plugins SynapCountJ y NucleusJ (que no utilizan directamente Kenzo sino un algoritmo de Fiji/ImageJ para calcular directamente el número de componentes conexas). Además, para el desarrollo del plugin NeuronPersistentJ se ha utilizado el nuevo módulo para Kenzo que permite el cálculo de homología persistente \cite{RHRS13}.

\subsection{Formalización}

 En el análisis de imágenes biomédicas es necesario el uso de software fiable; es decir, programas cuyos resultados deberían ser siempre correctos. Existen diversas técnicas para verificar formalmente la corrección de software (por ejemplo, con demostración automatizada por medio de \emph{model checking} y SAT o SMT \emph{solvers}), en nuestro caso hemos utilizado un método conocido como demostración asistida por ordenador.  Brevemente, este método consiste en utilizar un asistente para la demostración (en nuestro caso Coq \cite{Coq}) para comprobar que todos los pasos dados en una demostración son correctos; para una descripción detallada véase \cite{AD12}.

En nuestro trabajo, enmarcado en el proyecto ForMath \cite{Formath}, nos hemos centrado principalmente en la verificación de los algoritmos provenientes de la Topología Algebraica que utilizamos para analizar imágenes biomédicas. En \cite{CTIC12} presentamos la verificación de los programas necesarios para el cálculo de los grupos de homología de imágenes digitales. En la misma línea, en \cite{TOCL} introdujimos la formalización de los programas para el cálculo de la homología persistente de una pila de imágenes.

Los trabajos presentados en \cite{CTIC12} y \cite{TOCL} nos permiten realizar cálculos verificados; sin embargo, dichos programas no son capaces de trabajar con imágenes biomédicas debido a su tamaño. Conviene notar que las capacidades de cálculo de los asistentes para la demostración son limitadas, ya que su objetivo principal es la demostración de teoremas y no el cálculo. Para resolver el problema de eficiencia, en \cite{CICM12} presentamos la formalización de los algoritmos que utilizan campos vectoriales discretos, introducidos en \cite{RS10} (véase la sección 4 para un ejemplo de esta técnica). Estos programas nos permiten reducir la cantidad de información de las imágenes pero preservando sus propiedades topológicas. De este modo, combinando los trabajos \cite{CTIC12} y \cite{CICM12} se puede calcular de manera certificada grupos de homología de imágenes biomédicas.

Finalmente en \cite{CICM13} comenzamos el estudio de la corrección de los programas que implementan técnicas estándar para procesar imágenes (filtrados, eliminación de ruido). Esta línea de trabajo resulta especialmente interesante ya que algunos de los programas que han sido verificados son utilizados en una gran cantidad de plugins implementados en Fiji.

\section{Validación}

Los programas desarrollados para Fiji/ImageJ tratan de cubrir las necesidades del grupo de Plasticidad Sináptica Estructural. Para validar el funcionamiento de los plugins, se considera una batería de imágenes sobre las que los expertos realizan un procesamiento manual, y se comparan los resultados con los del proceso automático comprobando si la desviación entre ambos métodos es aceptable para los científicos experimentales. En algunos casos, si los resultados no son satisfactorios, se modifica el plugin siguiendo las observaciones de los biólogos para añadir más funcionalidad o ajustar algunos parámetros hasta lograr los resultados esperados.

Como ya se ha comentado anteriormente, en los procesos de recuento (y localización) manuales hay una componente subjetiva. Por ejemplo, a la hora de contar el número de núcleos de neuronas que aparecen en una imagen puede haber casos en los que el ojo humano no es capaz de apreciar con total seguridad si una de las componentes presentes en la imagen corresponde a una neurona o no. La experiencia muestra que algunos expertos cuentan \emph{de más} y otros \emph{de menos}. Sin embargo hay que destacar que el objetivo principal de los recuentos suele ser la comparación de resultados antes y después de realizar algún experimento (por ejemplo, en SynapCountJ el objetivo es estudiar la variación de la \emph{densidad sináptica}, y no tanto el número exacto de contactos sinápticos), por tanto lo que importa realmente es que en ambos casos el recuento se realice con los mismos criterios, ya sean más o menos \emph{exigentes}. Por esta misma razón, para el recuento automático realizado por los plugins no se pretende conseguir que el programa detecte todas las componentes (núcleos, espinas, etc.) presentes; se acepta una cierta desviación respecto al método manual siempre que el programa se comporte de forma coherente a la hora de comparar los resultados antes y después de realizar el experimento.

\begin{figure}
 \centering
\begin{tikzpicture}
\draw (0,0) node{\includegraphics[scale=.06]{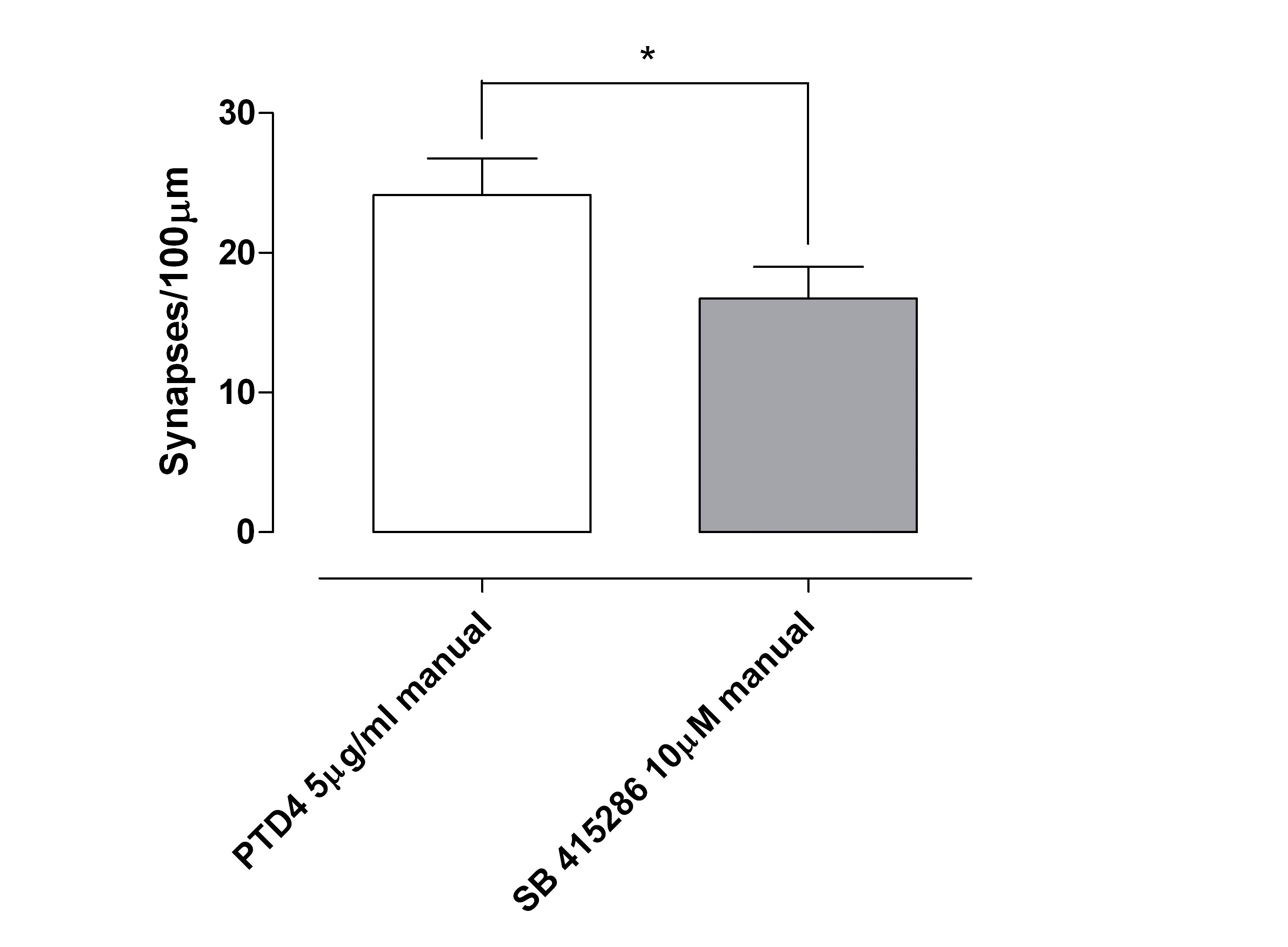}};
\draw (5,0) node{\includegraphics[scale=.06]{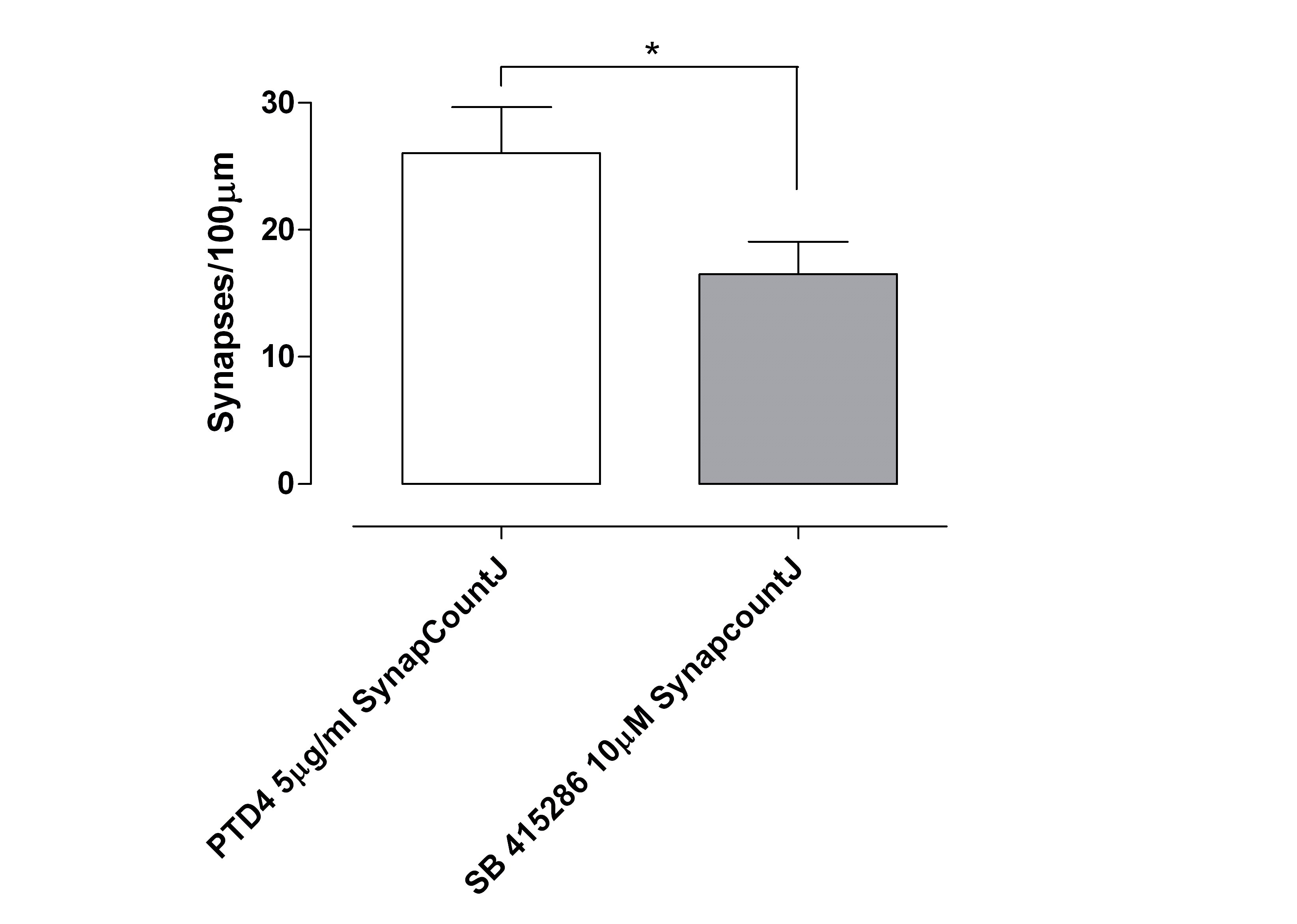}};
\end{tikzpicture}
 \caption{Datos de validación del plugin SynapCountJ}
 \label{fig:validacion}
\end{figure}

Por ejemplo, para validar el plugin SynapCountJ se realizó un estudio comparativo en el que se analizaron 13 fotografías individuales a neuronas de cultivos de 12 días (es decir, en el laboratorio se preparó una muestra de neuronas dejando pasar 12 días antes de ser estudiada). En los experimentos biológicos se suelen estudiar dos muestras: una en la que se ha aplicado la sustancia a estudiar (lo que se denomina abreviadamente ``tratamiento'') y otra en la que las medidas se realizan en el medio natural sin haber intervenido de ningún modo (``control''). De esta manera se puede saber si el tratamiento aplicado tiene algún efecto sobre la muestra. En el caso que nos ocupa, se trata de saber si el tratamiento tiene un efecto potenciador o inhibidor de la densidad sináptica. En la figura \ref{fig:validacion} se puede observar que al realizar la identificación y el recuento de manera manual se obtiene una media de 24.12 sinapsis en el control (PTD4 $5\mu g/ml$) y 16.74 sinapsis en el tratamiento (SB 415286 $10 \mu M$). Así mismo los resultados obtenidos mediante el \emph{plugin} son similares: se localizaron 26.03 sinapsis de media en el control y 16.50 sinapsis en el tratamiento. A pesar de las diferencias en el recuento, el porcentaje de inhibición medido con ambos procedimientos es el mismo, un $30.6\%$ de manera manual y un $36\%$ de modo automático. Hay que señalar que en este caso la densidad sináptica disminuye en el tratamiento frente al control  debido a que el estudio se ha realizado en cultivos de 12 días. Se ha constatado que para este tiempo de cultivo, el tratamiento tiene un efecto inhibidor de la densidad sináptica. Sin embargo en cultivos de 21 días el estudio muestra que la densidad sináptica puede aumentar aproximadamente un $60\%$, teniendo un efecto potenciador (véase \cite{MCMRH11}).

Respecto a los tiempos de ejecución, podemos indicar a modo de ejemplo que el procesado manual del cálculo del número de sinapsis de una imagen de una neurona puede requerir entre $30$ y $40$ minutos (y este tiempo puede aumentar a medida que el experto analiza varias imágenes), mientras que el cálculo de SynapCountJ es prácticamente inmediato. En el caso del recuento del número de neuronas, el experto necesita alrededor de una hora mientras que la ejecución del plugin NucleusJ tarda únicamente entre 1 y 2 minutos (incluyendo la especificación de los parámetros).

\section{Pragmática}
\label{sec:pragmatica}

Comentamos en esta sección algunos aspectos prácticos, relacionados con el tratamiento de imágenes digitales, que también han tenido importancia en el desarrollo de nuestro trabajo.

El primer aspecto que se debe tener en cuenta es el tamaño de las imágenes a tratar, que origina problemas en el cálculo (por ejemplo, de los grupos de homología), y especialmente en el proceso de  formalización ya que las capacidades de cálculo de los asistentes para la demostración son limitadas. Para solucionar este problema hemos utilizado la técnica de reducción de imágenes digitales presentada en la sección~\ref{sec:algoritmos} basada en la noción de \emph{campo vectorial discreto}.

Otro problema que nos hemos encontrado, habitual en el tratamiento de imágenes digitales, es el problema del \emph{ruido}: en las imágenes pueden observarse elementos que no se corresponden con la realidad y que aparecen de forma aleatoria debido al funcionamiento del microscopio. Para eliminar este efecto inevitable e indeseado se aplican diversas técnicas de preprocesamiento como el \emph{filtro de la mediana} o la \emph{deconvolución}.

Estas técnicas de preprocesamiento también son utilizadas para tratar de solucionar el problema de la \emph{Topología sin continuidad}, comentado en la sección \ref{sec:conceptos_mates}. En concreto, en todos los plugins desarrollados, y como paso previo a las técnicas topo-geométricas utilizadas posteriormente, se hace necesaria la aplicación de un \emph{threshold} o \emph{umbral} que haga más continua la imagen. La elección de un umbral adecuado es uno de los problemas más importantes que hemos debido afrontar, ya que dependiendo de las imágenes tratadas la elección adecuada puede variar mucho.

Por último, hay que señalar que para el procesamiento topológico (grupos de homología, homología persistente, persistencia zigzag) y en la formalización se debe trabajar con imágenes en blanco y negro. Esto hace necesario un procesamiento previo de binarización de las imágenes en el que de nuevo es necesario elegir un umbral adecuado.

En resumen, para obtener resultados útiles para los biólogos, es necesario evitar apriorismos matemáticos o informáticos,  y ``mancharse las manos'' en complejos preprocesamientos ad-hoc, guiados por el conocimiento experto de los científicos experimentales. El trabajar cerca de los biólogos permite también modificar las condiciones en las que las imágenes son captadas. Así, inyectando uno u otro marcador en un cultivo, los científicos consiguen proporcionar imágenes que destacan aspectos que permiten limitar el preprocesamiento y hacer que nuestros algoritmos funcionen de modo más eficaz.

\section{Conclusiones}
\label{sec:conclusiones}

Para el diagnóstico de enfermedades neurológicas y el estudio de la efectividad de los distintos fármacos propuestos se debe llevar a cabo un proceso de determinación de los diferentes elementos estructurales de las neuronas presentes en una imagen obtenida por el microscopio.  Hasta ahora esta tarea se realizaba de forma manual, siendo un trabajo lento, tedioso y con cierto grado de subjetividad. Tratando de solucionar estos problemas, y en colaboración con el grupo de biólogos PSE, hemos desarrollado una serie de programas, por medio de plugins para Fiji/ImageJ, que permiten un tratamiento automático de las imágenes y  facilitan la preparación de mecanismos para el procesamiento masivo de muestras, imprescindibles para dar el paso del laboratorio a la industria farmacéutica.

Los programas desarrollados permiten realizar el recuento automático del número de sinapsis presentes en una neurona, el recuento y la localización del número de neuronas que aparecen  en una imagen de cultivo, y la localización de una neurona a partir de una pila de imágenes correspondientes a distintas alturas. También se está desarrollando actualmente un nuevo programa para la localización de neuronas en una imagen de cultivo. Nuestras aplicaciones hacen uso de distintas ideas matemáticas como conexión, continuidad, distancia o árboles. La Topología Algebraica también tiene un papel fundamental a través de las nociones de grupos de homología, homología persistente y persistencia zigzag.


\begin{thebibliography}{10}

\bibitem{Formath}
{ForMath: Formalisation of Mathematics},
 \url{http://wiki.portal.chalmers.se/cse/pmwiki.php/ForMath/ForMath}.

\bibitem{AD12}
\textsc{J.~Aransay y C.~Domínguez},
 Demostración asistida por ordenador,
\textit{Gaceta de la RSME} \textbf{15(1)}(2012), 75--104.

\bibitem{Coq}
\textsc{Y.~Bertot y P.~Casteran},
\textit{Interactive Theorem Proving and Program Development, Coq’Art:
  the Calculus of Constructions},
Springer-Verlag, 2004.

\bibitem{CS10}
\textsc{G.~Carlsson y V.~de~Silva},
 Zigzag persistence,
\textit{Foundations of Computational Mathematics} \textbf{10} (2010), 367--405.

\bibitem{Kenzo}
\textsc{X.~Dousson, J.~Rubio, F.~Sergeraert e Y.~Siret},
 The {Kenzo} program,
Institut Fourier, Grenoble, 1999,
 \url{http://www-fourier.ujf-grenoble.fr/~sergerar/Kenzo/}.

\bibitem{EH08}
\textsc{H.~Edelsbrunner y J.~Harer},
Persistent homology - a survey,
\textit{Surveys on Discrete and Computational Geometry. Twenty Years
  Later. Contemporary Mathematics}  \textbf{453} (2008), 257--282.

\bibitem{For98}
\textsc{R.~Forman},
Morse theory for cell complexes,
\textit{ Advances in Mathematics}  \textbf{134} (1998), 90--145.

\bibitem{TOCL}
\textsc{J.~Heras, T.~Coquand, A.~Mörtbert y V.~Siles},
 {Computing Persistent Homology within Coq/SSReflect},
será publicado en el \textit{ ACM Transactions on Computational Logic},
  2013.

\bibitem{CTIC12}
\textsc{J.~Heras, M.~Dénès, G.~Mata, A.~Mörtberg, M.~Poza y V.~Siles},
Towards a certified computation of homology groups for digital
  images,
\textit{Proceedings CTIC 2012, Lecture Notes in Computer Science}
  \textbf{7309} (2012), 49--57.

\bibitem{CICM13}
\textsc{J.~Heras, G.~Mata, A.~Romero, J.~Rubio y R.~Sáenz},
{Verifying a platform for digital imaging: a multi-tool strategy},
\textit{ Proceedings CICM 2013, Lecture Notes in Computer Science}
  \textbf{7961} (2013), 66--81.

\bibitem{HPR12}
\textsc{J.~Heras, V.~Pascual y J.~Rubio},
A certified module to study digital images with the {K}enzo system,
\textit{Lecture Notes in Computer Science} \textbf{6927} (2012), 113--120.

\bibitem{CICM12}
\textsc{J.~Heras, M.~Poza y J.~Rubio},
{Verifying an algorithm computing Discrete Vector Fields for digital
  imaging},
\textit{ Proceedings CICM 2012, Lecture Notes in Computer Science}
  \textbf{7362} (2012), 215--229.

\bibitem{SynapCountJ}
\textsc{G.~Mata},
{SynapCountJ},
 Universidad de La Rioja, 2011,
  \url{http://imagejdocu.tudor.lu/doku.php?id=plugin:utilities:synapsescountj:start}.

\bibitem{NeuronPersistentJ}
\textsc{G.~Mata},
{NeuronPersistentJ},
Universidad de La Rioja, 2012,
  \url{http://imagejdocu.tudor.lu/doku.php?id=plugin:utilities:neuronpersistentj:start}.

\bibitem{NucleusJ}
\textsc{G.~Mata},
{NucleusJ},
Universidad de La Rioja, 2013,
\url{http://spineup.jimdo.com/downloads/}.

\bibitem{MCMRH11}
\textsc{G.~Mata, G.~Cuesto, M.~Morales, J.~Rubio y J.~Heras},
SynapCountJ: un software para el estudio de la densidad sináptica,
Presentado en XIV Congreso de la Sociedad Española de Neurociencia
  (SENC 2011), 2011.

\bibitem{NeuronJ}
\textsc{E.~Meijering, M.~Jacob, J.C. Sarria, H.~Hirling y M.~Unser},
Design and validation of a tool for neurite tracing and analysis in
  fluorescence microscopy images,
\textit{Cytometry Part A} \textbf{58(2)} (2004), 167--176.

\bibitem{MW10}
\textsc{R.~Mikhailov y J.~Wu},
On homotopy groups of the suspended classifying spaces,
\textit{ Algebraic and Geometric Topology} \textbf{10} (2010), 565--625.

\bibitem{ImageJ}
\textsc{W.~S. Rasband},
ImageJ: Image Processing and Analysis in Java,
Technical report, U. S. National Institutes of Health, Bethesda,
  Maryland, USA, 1997--2012.

\bibitem{RHRS13}
\textsc{A.~Romero, J.~Heras, J.~Rubio y F.~Sergeraert},
Defining and computing persistent $\mathbb{Z}$-homology in the
  general case,
Preprint, 2013.

\bibitem{RR12}
\textsc{A.~Romero y J.~Rubio},
Homotopy groups of suspended classifying spaces: an experimental
  approach,
\textit{ Mathematics of Computation} \textbf{82} (2013), 2237--2244.

\bibitem{RS10}
\textsc{A.~Romero y F.~Sergeraert},
Discrete {V}ector {F}ields and fundamental {A}lgebraic {T}opology,
Preprint, \url{http://arxiv.org/abs/1005.5685v1}, 2010.

\bibitem{Fiji}
\textsc{J.~Schindelin et~al},
Fiji: an open-source platform for biological-image analysis,
\textit{Nature Methods} \textbf{9(7)} (2012), 676--682.

\end{thebibliography}
\end{document}